\documentclass[twocolumn, usenames, dvipsnames]{aastex62}

\usepackage{graphicx}	          % Including figure files
\usepackage{amsmath}	          % Advanced maths commands
\usepackage{upgreek}            % uptau
\usepackage{amssymb}	          % Extra maths symbols
\usepackage{bm}		              % Bold maths symbols, including upright Greek
\usepackage{CJKutf8}            % Chinese name
\usepackage{comment}            % Comment big block
\usepackage{tabularx}           % Newer table package
\usepackage{threeparttable}     % Table notes smart width
\usepackage{booktabs}           % Better \hline
\usepackage{soul}               % better underline
\usepackage{multirow}           % to enable multirow cells in tables
\usepackage[T1]{fontenc}        % for special char in Drazkowska2014
\usepackage{microtype}
\usepackage{xcolor}
\DisableLigatures[f]{encoding = *, family = *} %% <- only disables f-ligatures
\newcommand{\se}{\mathrm{e}} % used in scientific notation
\def\num#1{\numx#1}\def\numx#1e#2{{#1}\mathrm{e}{#2}}
\newcommand{\mm}[1]{\mathrm{#1}}
\newcommand{\fracd}[2]{\frac{d{#1}}{d{#2}}}

\newcommand{\fracp}[2]{\frac{\partial{#1}}{\partial{#2}}}
\newcommand{\fracpp}[2]{\frac{\partial^2{#1}}{\partial{#2}^2}}

\makeatletter
\newcommand{\pushright}[1]{\ifmeasuring@#1\else\omit\hfill$\displaystyle#1$\fi\ignorespaces}
\newcommand{\pushleft}[1]{\ifmeasuring@#1\else\omit$\displaystyle#1$\hfill\fi\ignorespaces}
\makeatother

\received{--}
\revised{--}
\accepted{--}
\submitjournal{ApJ}

\shorttitle{PLanetesimal ANalyzer}
\shortauthors{Li et al.}
\begin{document}

\title{Demographics of Planetesimals Formed by the Streaming Instability}
%\title{A Statistically Robust Census of Planetesimals Formed by the Streaming Instability}
%\title{On the Initial Mass Distribution of Planetesimals Formed by the Streaming Instability}

\correspondingauthor{Rixin Li}
\email{rixin@email.arizona.edu}

\author[0000-0001-9222-4367]{Rixin Li
\begin{CJK*}{UTF8}{gkai}
  (李日新)
\end{CJK*}}
\affil{Steward Observatory \& Department of Astronomy, University of Arizona, \\
 933 N Cherry Ave, Tucson, AZ 85721, USA}

\author[0000-0002-3644-8726]{Andrew N. Youdin}
\affiliation{Steward Observatory \& Department of Astronomy, University of Arizona, \\
933 N Cherry Ave, Tucson, AZ 85721, USA}

\author[0000-0002-3771-8054]{Jacob B. Simon}
\affiliation{Department of Physics and Astronomy, Iowa State University, Ames, IA, 50010, USA}
\affiliation{JILA, University of Colorado and NIST, 440 UCB, Boulder, CO 80309-0440}
\affiliation{Department of Space Studies, Southwest Research Institute, Boulder, CO 80302}

%%%%%%%%%%%%%%%%%%%%%%%%%%%%%%%%%%%%%%%%%%%%%%%%%%%%%%%%%%%%%%%%%%%%%%%%%%%%%%%%
\begin{abstract}
%% Context
The streaming instability (SI) is a mechanism to aerodynamically concentrate solids in protoplanetary disks and facilitate the formation of planetesimals.   Recent numerical modeling efforts have demonstrated the increasing complexity of the initial mass distribution of planetesimals.
%% Aims
To better constrain this distribution, we conduct SI simulations including the self-gravity with hitherto the highest resolution.  To subsequently identify all of the self-bound clumps, we develop a new clump-finding tool, PLanetesimal ANalyzer (\texttt{PLAN}).  
%% Methods
We then apply a maximum likelihood estimator to fit a suite of parameterized models with different levels of complexity to the simulated mass distribution.  To determine which models are best-fitting and statistically robust, we apply three model selection criteria with different complexity penalties.
% Conclusions
We find that the initial mass distribution of clumps is not universal regarding both the functional forms and parameter values.  Our model selection criteria prefer models different from those previously considered in the literature.  Fits to multi-segment power law models break to a steeper distribution above masses close to 100 km collapsed planetesimals, similar to observed Kuiper Belt size distributions. We find evidence for a turnover in the low mass end of the planetesimal mass distribution in our high resolution run. Such a turnover is expected for gravitational collapse, but had not previously been reported.
\end{abstract}


%%%%%%%%%%%%%%%%%%%%%%%%%%%%%%%%%%%%%%%%%%%%%%%%%%%%%%%%%%%%%%%%%%%%%%%%%%%%%%%%
\section{Introduction}
\label{sec:intro}

% outline 1 - planetesimal formation and the SI
An indispensable step in planet formation is to build planetesimals---super-kilometer objects bound by self-gravity---in protoplanetary disks \citep{Chiang2010, Johansen2014}.  One of the compelling pathways to planetesimal formation is the efficient concentration of solids by the streaming instability (SI) followed by their gravitational collapse \citep{Youdin2005, Johansen2007a}.  

% outline 2 - more explanations on the SI
The SI is an aerodynamic instability arising from the relative drift and the mutual drag forces between gas and solids \citep{Youdin2005}.  The SI is  one example of a broader class of drag instabilities in protoplanetary disks \citep{Goodman2000,Lin2017, Squire2018}.  

Strong particle clumping can be induced by the SI under the right conditions, that is, when the midplane dust-to-gas volume density ratio exceeds unity, which further depends on the particle size, dust-to-gas surface density ratio (often referred as metallicity), and gas pressure gradient \citep{JY2007, Bai2010a, Bai2010b, Carrera2015, Yang2017}.  In a typical smooth disk with cm-sized pebbles, slightly super solar metallicity is required to trigger the strong SI \citep{Johansen2009a}.  Higher metallicity may be reached by gas removal (e.g., via photoevoporation) or dust pile-up (e.g., at pressure bumps, snow lines), where smaller solids can also lead to strong particle clumping \citep{Carrera2017, Drazkowska2016, Drazkowska2017, Schoonenberg2017}.  Previous studies have also shown that high-resolution SI simulations with self-gravity can produce a broad initial top-heavy mass distribution of planetesimals \citep{Johansen2015, Simon2016, Simon2017}.  

Recently, \citet{Yang2018} suggest that the midplane dust-to-gas density ratio exceeding order unity may not be necessary for the strong SI when the radial diffusion of particles driven by turbulence near the midplane is weak.  \citet{Lin2019} find that particle back-reaction on to the gas can lead to self-sustained dust sedimentation against the vertical shear instability turbulence at a modest super solar metallicity, which may trigger the SI.  Furthermore, \citet{Krapp2019} investigate the linear growth of the SI with multiple dust species for the first time\deleted{ and find a properly resolved particle-size distribution can significantly affect the linear phase of the SI, which merit more studies and stratified simulations with sufficient dust species }\added{, showing how a properly resolved particle-size distribution affects the linear phase of the SI. Their study motivates more  detailed simulations of the non-linear phase of the multi-species streaming instability as well, which would extend previous studies}\citep{Bai2010a, Schaffer2018}.

% outline 2.5 - observation motivations
Observations of asteroids and Trans-Neptunian objects support models in which planetesimals were born big \citep{Morbidelli2009}, with evidence of a drop-off in planetesimal numbers below $\sim$1--50 kilometers, depending on the population \citep{Delbo2017, Singer2019}.  \citet{Nesvorny2019} find that SI simulations correctly predict the primarily prograde mutual inclinations of the abundant binaries in the Cold Classical Kuiper Belt \citep{Grundy2019}.  Further studies on the demographics of planetesimals formed via the SI, and by other mechanisms, offer the promise of more detailed observational comparisons and tests.

% outline 3 - the issue we want to address in this work
Quantifying the mass distribution of planetesimals formed by the SI is of significant interest.  Due to the high computational cost of SI simulations, a parameterized mass function can be used as the input for global studies of disk evolution and planet formation \citep{Drazkowska2014}.  Furthermore, the shape of the mass distribution offers insights to the physical processes of particle clumping and gravitational collapse.

Previous work has fit the mass distribution to a simple power law \citep{Simon2016, Simon2017} or to a power law with an exponential cutoff or truncation \citep{Schafer2017, Abod2019}.  These work suggested that the initial planetesimal mass function might be near-universal.   However, it is not trivial to determine the best parameterization of the initial planetesimal mass function in SI simulations.  Moreover, it is not clear if a single functional form can describe planetesimal formation with different physical conditions, i.e. simulation parameters.

% outline 4 - our goal in this paper
Motivated by these issues, our goal in this paper is to better understand and constrain the broad initial mass distribution of planetesimals with robust statistical analyses.   This work will fit many different parameterizations to simulated planetesimal mass distributions.  To determine which models best describe the data, model selection techniques weigh the goodness of fit against a complexity penalty, intended to avoid the overfitting of data features that might be spurious.  Since there is no universal agreement on complexity penalties either, we apply different model selection techniques, including a bootstrap method that we developed independently.  

% outline 5 - paper structures, may be modified later
The paper is organized as follows.  In Section \ref{sec:method}, we begin with an overview of the numerical models and our simulations. Section \ref{subsec:plan} then introduce our newly-developed clump-finding tool, \texttt{PLAN}. Section \ref{sec:fitting} lays out all the statistical models and our fitting procedure as well as the model selection criteria.  In Section \ref{sec:results}, we show the fitting results and the model selection results.  Section \ref{sec:final} discusses the implications of our statistical understanding, with a summary and conclusions in the end.

%%%%%%%%%%%%%%%%%%%%%%%%%%%%%%%%%%%%%%%%%%%%%%%%%%%%%%%%%%%%%%%%%%%%%%%%%%%%%%%%
\begin{deluxetable*}{c|c|c|c|c|c|c|c}
  \tablecaption{Simulation Parameters\label{tab:setup}}
  \tablecolumns{7}
  %\tablewidth{0.0\linewidth}
  \tablehead{
    \colhead{Run} &
    \colhead{Domain Size} &
    \colhead{Number of Cells} &
    \colhead{$N_{\rm par}$\tablenotemark{$*$}} &
    \colhead{$\uptau_{\rm s}$} &
    \colhead{$Z$} &
    \colhead{$t_{0}$\tablenotemark{$\dag$}} &
    \colhead{$N_{\rm tot}$\tablenotemark{$\ddag$}} \\
    \colhead{} &
    \colhead{$(L_X\times L_Y\times L_Z)H^3$} &
    \colhead{$N_X\times N_Y\times N_Z$} &
    \colhead{} &
    \colhead{} &
    \colhead{} &
    \colhead{$(\Omega_0^{-1})$} &
    \colhead{}
  }
  \startdata
  \hline\hline
    I  &  $0.1\times 0.1\times 0.2$  &  $512\times 512\times 1024$  &  $134,217,728$  &  $2.0$  &  $0.1$   &  $36.0$  &  $284$\\
   %B22  &  $0.2\times 0.2\times 0.2$  &  $512\times 512\times 512$   &  $153,600,000$  &  $2.0$  &  $0.1$   &  $37.0$  \\
   II  &  $0.2\times 0.2\times 0.2$  &  $512\times 512\times 512$   &  $153,600,000$  &  $0.3$  &  $0.02$  &  $110.0$  & $174$
  \enddata
  \tablecomments{For all runs: the radial pressure gradient term is $\Pi = 0.05$ and the particle self-gravity strength is $\tilde{G} = 0.05$. }
  \tablenotetext{$*$}{The number of particles.  For reference, $2^{27}=512^3=134,217,728$.}
  \tablenotetext{$\dag$}{The time when the particle self-graivty is switched on.}
  \tablenotetext{$\ddag$}{The number of clumps identified by \texttt{PLAN} at the snapshot where we perform analyses and fitting in Section \ref{sec:results}.}
\end{deluxetable*}

\begin{figure*}
  \centering
  \includegraphics[width=\linewidth]{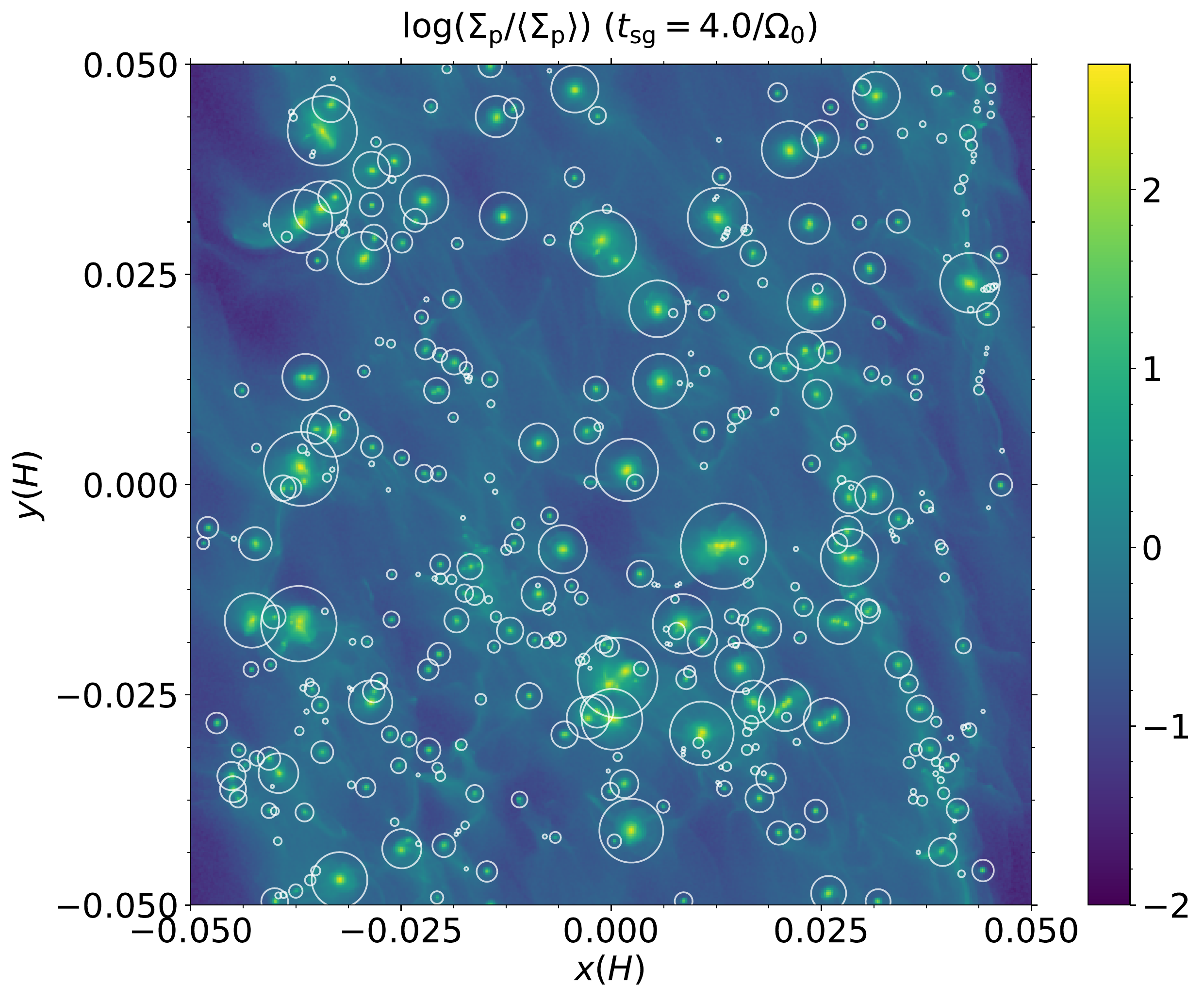}
  \caption{A snapshot of the solid surface density ($\Sigma_{\rm p}$) from the simulation Run I.  This snapshot is $4/\Omega_0$ after the particle self-gravity has been switched on, where self-bound clumps have already formed from collapse.  All of the clumps identified by \texttt{PLAN} are marked by white circles that illustrate their Hill spheres. \label{fig:snapshot}}
\end{figure*}

%%%%%%%%%%%%%%%%%%%%%%%%%%%%%%%%%%%%%%%%%%%%%%%%%%%%%%%%%%%%%%%%%%%%%%%%%%%%%%%%
\section{Method}
\label{sec:method}

To simulate the formation of planetesimals, we use the \texttt{ATHENA} code with a similar setup to \citet{Simon2017}.  In Section \ref{subsec:athena}, we briefly introduce the numerical methods employed in \texttt{ATHENA} for modeling the coupled dynamics of gas and particles, including the self-gravity of solids, in a protoplanetary disk \citep[see][for more details]{Bai2010, Simon2016}.  Section \ref{subsec:setup} then summarizes the numerical setup and parameters used in our simulations. Section \ref{subsec:plan} explains how \texttt{PLAN} identifies and characterizes all the self-bound clumps in the output particle data.

%%%%%%%%%%%%%%%%%%%%%%%%%%%%%%%%%%%%%%%%%%%%%%%%%%%%%%%%%%%%%%%%%%%%%%%%%%%%%%%%
\subsection{Planetesimal Formation Simulations}
\label{subsec:athena}

% local box approximation
We use \texttt{ATHENA} to simulate a small three-dimensional vertically-stratified patch of the protoplanetary disk with the local shearing box approximation \citep{Stone2008, Stone2010, Hawley1995}.  This approximation---which is justified by the small length scales of the SI compared to the radial position in the disk---maps the global disk geometry $(R, \phi, z')$ onto a local Cartesian coordinate system $(x, y, z)$ \citep{Goldreich1965}.  The local box is centered at a fiducial disk radius ($R_0$) in the midplane, where $(x, y, z) \equiv (R-R_0, R_0 \phi, z')$, where the Keplerian frequency and velocity are $\Omega_0$ and $v_{\rm K} = \Omega_0 R_0$, respectively.

% equations of motion
In this non-inertial computational domain, \texttt{ATHENA} solves the equations of gas dynamics and the equation of motion for each particle (indexed by $i$)
\begin{align}
  \fracp{\rho_{\rm g}}{t} + \nabla \cdot (\rho_{\rm g} \bm{u}) &= 0, \label{eq:gascon}\\
\begin{split}\label{eq:gasmom}
  \fracp{(\rho_{\rm g} \bm{u})}{t} + \nabla\cdot(\rho_{\rm g} \bm{u}\bm{u} + P\bm{I}) &=\\
  \rho_{\rm g} \biggl[ 2\bm{u}\times\bm{\Omega}_0 + 3{\Omega}_0^2 \bm{x} &- {\Omega}_0^2 \bm{z} \biggr] + \rho_{\rm p} \frac{\bar{\bm{v}} - \bm{u}}{t_\mm{stop}},
\end{split} \\
\begin{split}\label{eq:ithpar}
  \fracd{\bm{v}_i}{t} = 2\bm{v}_i\times\bm{\Omega}_0+3{\Omega}_0^2 \bm{x}_i &- {\Omega}_0^2 \bm{z}_i \\
  -\frac{\bm{v}_i - \bm{u}}{t_\mm{stop, i}} &- \nabla\Phi_{\rm sg} - 2\eta v_{\rm K} \Omega_0 \hat{x},
\end{split}
\end{align}
where $\rho_{\rm g}$, $\bm{u}$ and $P$ are density, velocity and pressure of gas, $\bm{I}$ is the identity matrix, $\bm{\Omega}_0 = \Omega_0 \hat{z}$, $\rho_{\rm p}$ and $\bar{\bm{v}}$ are the average density and velocity of the particles in a hydrodynamic grid cell, $t_{\rm stop}$ is the dimensional stopping time, $\bm{v}_i$ is the velocity of the $i$-th particle, $\Phi_{\rm sg}$ is the potential field of the self-gravity of solids, and $\eta$ denotes the relative difference between the gas orbital velocity and the Keplerian velocity due to the radial pressure gradient in the disk.

Our model calculates the Coriolis forces, radial and vertical tidal gravity, and the particle feedback exerted on the gas, as in the right hand side of Eq. \ref{eq:gasmom}.  The equation of state for the gas is assumed to be isothermal, $P = c_{\rm s}^2 \rho_{\rm g}$, where the constant $c_{\rm s}$ is the isothermal sound speed.  We neglect the self-gravity of the gas because the gas density fluctuations are relatively negligible.

For solids, \texttt{ATHENA} adopts the super-particle treatment, where each particle in our simulations statistically represents a large number of pebbles in terms of mass.  The acceleration of each particle is governed by Eq. \ref{eq:ithpar} with the Coriolis and tidal forces (similar to those in Eq. \ref{eq:gasmom}), and also the gas drag as well as the force due to particle self-gravity.  The gravitational potential field, $\Phi_{\rm sg}$, is obtained by solving Poisson's equation
\begin{equation}
  \nabla^2\Phi_{\rm sg} = 4 \pi G \rho_{\rm p}
\end{equation}
with the Fast Fourier Transform (FFT) method of \citet{Simon2016}, where $G$ is the gravitational constant.  The accuracy of particle self-gravity from such a method depends on the grid resolution.  The last source term in Eq. \ref{eq:ithpar}, $- 2\eta v_{\rm K} \Omega_0 \hat{x}$, is a constant radial force in \texttt{ATHENA} to implement the effective global radial pressure gradient under the restrictions of the local model.

The radial and azimuthal boundary conditions (BCs) for our model are the standard shearing-periodic BCs \citep{Stone2010}.  In the vertical direction, we use a modified outflow BCs that extrapolates the gas density into the ghost zones exponentially and prohibits any gas inflow \citep{Simon2011, Li2018}.  These vertical BCs maintain hydrostatic equilibrium and reduce artificial gas motions at vertical boundaries, which is beneficial for simulations in short boxes.  Furthermore, the total gas mass is renormalized to compensate the gas outflow at each time step to ensure the mass conservation. 

The physical behavior of our simulations are dominated by four key dimensionless parameters.  The SI is characterized by the first three of them: the dimensionless particle stopping time
\begin{equation}
  \uptau_{\rm s} = \Omega_0 t_{\rm stop},
\end{equation}
which represents the ratio of a particle’s aerodynamic ($t_{\rm stop}$) and orbital ($\Omega_0^{-1}$) timescales, increases with a particle’s size, and decreases with the local gas density; the surface density ratio between the solids ($\Sigma_{\rm p}$) and the gas ($\Sigma_{\rm g}$)
\begin{equation}
  Z = \frac{\Sigma_{\rm p}}{\Sigma_{\rm g}},
\end{equation}
which is sometimes called the total solid-to-gas mass ratio; the global radial pressure gradient parameter
\begin{equation}
  \Pi \equiv \frac{\eta v_{\rm K}}{c_{\rm s}} \equiv -\frac{1}{2}\frac{c_{\rm s}}{v_{\rm K}}\fracp{\ln{P}}{\ln{R}},
\end{equation}
which accounts for the strength of the headwind on the particles.  The fourth key parameter controls the relative strength of the particle self-gravity compared with the tidal shear
\begin{equation}
  \tilde{G} \equiv \frac{4 \pi G \rho_0}{\Omega_0^2} = \frac{4}{\sqrt{2\pi}Q},
\end{equation}
where $\rho_0$ is the midplane gas density, $Q$ is the Toomre's Q \citep{Toomre1964}.

The code units of our simulations are set to the natural units of the shearing box.  The density unit and the time unit are $\rho_0$ and $\Omega_0^{-1}$, respectively.  While $\rho_0$ and $\Omega_0$ are code units, their allowed physical values and thus the choice of disk model are constrained by the $\tilde{G}$ parameter.  The length unit is $H = c_{\rm s}/\Omega_0$, the vertical scale height of the gas.

\subsection{Numerical Setup}
\label{subsec:setup}

All of our simulations are initiated with Gaussian vertical density profiles for both the gas and particles
\begin{equation}
  \begin{aligned}
    \rho_{\rm g} &= \rho_0 \exp\left(\frac{-z^2}{2H^2}\right), \\
    \rho_{\rm p} &= \frac{\Sigma_{\rm p}}{\sqrt{2\pi}H_{\rm p}} \exp\left(\frac{-z^2}{2H_{\rm p}^2}\right),
  \end{aligned}
\end{equation}
where $H_{\rm p}$ is the particle scale height and is set to $0.02H$ in the beginning.  The choice of such an initial particle scale height matches our previous work in \citet{Simon2017}, which let particles naturally sediment to a pseudo-equilibrium scale height of the order of $0.01H$ \citep{Yang2014, Li2018}.  Both gas and particles are then initialized with the Nakagawa–Sekiya–Hayashi (NSH) equilibrium drift velocities \citep{Nakagawa1986}.

In this work, we fix $\Pi = 0.05$ and $\tilde{G} = 0.05$ ($Q\simeq 32$), which are typical values in protoplanetary disks.  Table \ref{tab:setup} lists other physical and numerical parameters for all two of our simulations.  Run I has $(\uptau_{\rm s}, Z) = (2.0, 0.1)$ and a higher resolution in a smaller domain (one grid cell $\Delta x$ is $H/5120$ wide, the highest resolution to date).  Run II has $(\uptau_{\rm s}, Z) = (0.3, 0.02)$ and a lower resolution in a larger domain.  The data of Run II are directly taken from \citet{Simon2017}, and our Run I is a higher resolution version of another simulation in \citet{Simon2017}.  The relation between particle size and  $\uptau_{\rm s}$ depends on uncertain properties of particles and gas disk.  The range of $\uptau_{\rm s}$ adopted here may correspond to solids with any size from millimeter to decimeter.  These physical parameters are chosen known to produce strong particle clumping that triggers gravitational collapse.  The resulting bound clumps are the subject of our statistical analyses below.

Following previous studies, we start our simulations first without particle self-gravity.  Only after the SI has fully developed and saturated, do we switch on the self-gravity.  This approach significantly reduces the computational expense and has little influence on the final properties of planetesimals \citep{Simon2016, Abod2019}.  For convenience, we define a self-gravity time
\begin{equation}
  t_{\rm sg} = t - t_0,
\end{equation}
where $t$ is the simulation time and $t_0$ denotes the time when the self-gravity is turned on (see the next to last column in Table \ref{tab:setup}).  Also, we present all planetesimal masses in units of the dimensional mass for a self-gravitating particle disk
\begin{equation}
    M_{\rm G} = \pi \left(\frac{\lambda_{\rm G}}{2}\right)^2\Sigma_{\rm p} = 4\pi^5\frac{G^2\Sigma_{\rm p}^3}{\Omega_0^4} =  \frac{\sqrt{2}}{2}\pi^\frac{9}{2}Z^3\tilde{G}^2 (\rho_0 H^3),
\end{equation}
where $\lambda_{\rm G}$ is the critical unstable wavelength from the standard Toomre dispersion relation.  

To translate $M_{\rm G}$ into a physical mass unit and planetesimal size, additional assumptions about the disk model are required.  For instance, if we assume a fiducial disk radius of $R_0 = 10$ au in the modified minimum mass solar nebula (MMSN) model of  \citet[adapted from \citealt{Hayashi1981}, hereafter CY10 model]{Chiang2010}, where $\Sigma_{\rm g}\propto R^{-3/2}$ and the temperature  $T\propto R^{-3/7}$, then our $\tilde{G}=0.05$ parameter implies the gas mass in the CY10 model is about half % = 0.538
the original MMSN values.  For these parameters, the CY10 model has $\Pi = 0.068$, slightly higher than $\Pi = 0.05$ in our simulations.  Nevertheless, a smaller $\Pi$ value might arise from a weak pressure bump, a common substructure in protoplanetary disks \citep{Pinilla2017, dullemond2018}.  Moreover, $\Pi$ values do not significantly affect the planetesimal mass distribution, as studied in \citet{Abod2019}.  The mass unit for Run I then equates to $M_{\rm G} = 1.82\times 10^{23}$ g $= 0.19\ M_{\rm Ceres}$.  With $1$ g cm$^{-3}$ as the mean density, the physical radius of a $1 M_{\rm G}$ is $\simeq 350$ km in this model.  For Run II, the same gas disk model and location gives $M_{\rm G} = 1.45\times 10^{21}$ g $= 0.0015\ M_{\rm Ceres}$, and a physical radius of $\simeq 70$ km.

Fig. \ref{fig:snapshot} shows a snapshot of the particle surface density from Run I at $t_{\rm sg} = 4/\Omega_0$, where solids already collapse into self-bound clumps.  In the following section, we describe how \texttt{PLAN} finds these clumps in detail.

%%%%%%%%%%%%%%%%%%%%%%%%%%%%%%%%%%%%%%%%%%%%%%%%%%%%%%%%%%%%%%%%%%%%%%%%%%%%%%%%

\begin{figure*}
  \centering
  \includegraphics[width=\linewidth]{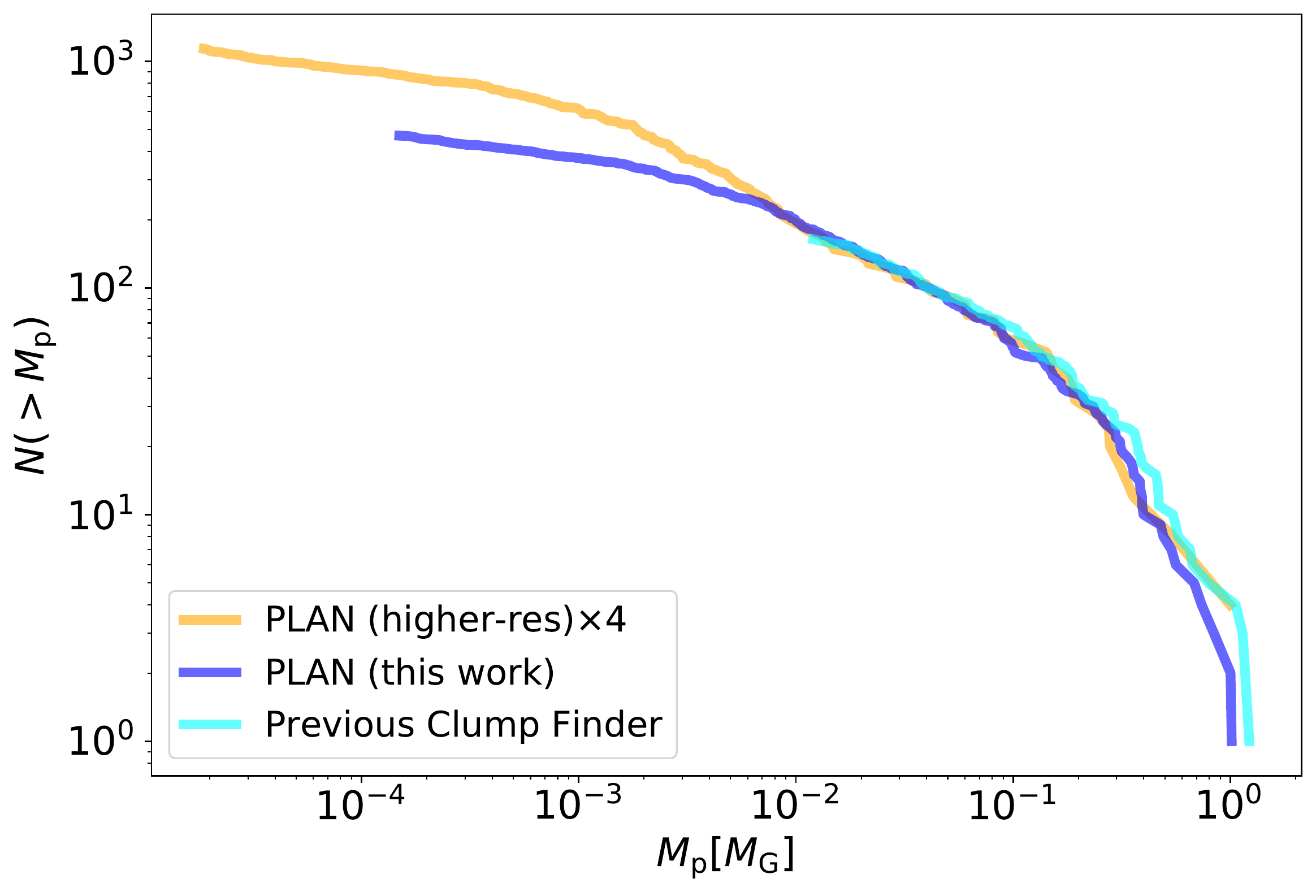}
  \caption{Cumulative number of bound, planetesimal-forming clumps above a given mass as measured in SI simulations with the same physical parameters as Run I.  The effects of both the clump finding algorithm -- PLAN (\textit{blue}) vs. a previous method (\textit{cyan}) -- and simulation resolution -- higher (\textit{orange}) vs. lower (\textit{blue}) -- are shown. 
  Both the \textit{cyan} and \textit{blue} curves analyze a simulation snapshot from \citet{Simon2017} which has half the resolution and twice the box size as our run I.  The \textit{orange} curve analyzes a  Run I snapshot, with planetesimal numbers augmented by a factor of 4 to compensate for the smaller surface area. 
  Comparing  \textit{blue} and \textit{cyan}, PLAN finds smaller planetesimals than the previous method, and also gives lower masses for the largest planetesimals, by differentiating vertically overlapping clumps (see text for more details).  Comparing \textit{orange} and \textit{blue}, higher numerical resolution extends the mass distribution to lower masses ($<10^{-4} M_{\rm G}$), amends results at intermediate masses ($<10^{-2} M_{\rm G}$) and agrees with lower resolution simulations at the high mass end ($>10^{-2} M_{\rm G}$). \label{fig:CDF}}
\end{figure*}

%%%%%%%%%%%%%%%%%%%%%%%%%%%%%%%%%%%%%%%%%%%%%%%%%%%%%%%%%%%%%%%%%%%%%%%%%%%%%%%%
\subsection{Clump-Finding with PLanetesimal ANalyzer}
\label{subsec:plan} % fixed

To identify and further characterize the properties of planetesimals produced in our simulations, we develop a new clump-finding tool, PLanetesimal ANalyzer (\texttt{PLAN}, \citet{PLAN}).  It is designed to work with the 3D particle output of \texttt{ATHENA} and find self-bound clumps robustly and efficiently.  \texttt{PLAN} is scalable to analyze billions of particles and many snapshots simultaneously because of its massively parallelized scheme written in C\nolinebreak\hspace{-.05em}\raisebox{.4ex}{\tiny\bf +}\nolinebreak\hspace{-.10em}\raisebox{.4ex}{\tiny\bf +} with OpenMP/MPI.

We now briefly present the workflow of \texttt{PLAN}.  The approach is based on the dark matter halo finder \texttt{HOP} developed by \citet{EH1998}, which is able to quickly group physically related particles.  \texttt{PLAN} first builds a memory-efficient linear Barnes-Hut tree representing all the particles in the Morton order \citep{BH1986}.  Each particle is then assigned a density computed from the nearest $N_{\rm den}$ particles ($N_{\rm den} = 64$ by default).  For particles with densities higher than a threshold, $\delta_{\rm outer} = 8\rho_0/\tilde{G}$, \texttt{PLAN} chains them up towards their densest neighbors recursively until a density peak is reached.  The value of $\delta_{\rm outer}$ is physically motivated to be slightly smaller than the Roche density ($9\rho_0/\tilde{G}$) such that PLAN can quickly find most relevant particles.  All the particle chains leading to the same density peak are combined into a group.  

\texttt{PLAN} then merges those groups by examining their boundaries to construct a list of bound clumps.  Based on the total kinematic and gravitational energies, deeply intersected groups are merged if bound.  However, two particle groups with a saddle point less dense than $\delta_{\rm saddle}=2.5\delta_{\rm outer}$ remain separated \citep{EH1998}.  Next, \texttt{PLAN} goes through each group---or raw clump---to unbind any contamination (i.e., passing-by and not bound) particles and gather possibly unidentified member particles within its Hill sphere.  After discarding those clumps (if any) with Hill radii ($R_{\rm Hill}$) smaller than one hydrodynamic grid cell ($\Delta x$) or density peaks less than $\delta_{\rm peak}=3\delta_{\rm outer}$, \texttt{PLAN} outputs the final list of clumps with their physical properties derived from particles.

Most clumps in our high-resolution simulations are highly-concentrated, where particles often collapse into regions much smaller than the Hill radius (see Fig. \ref{fig:snapshot}).  For small clumps, these regions are comparable to one cell size.  The particle-mesh method of calculating self-gravity does not resolve scales below $\Delta x$, which is a primary motivation for our high-resolution simulation and the reason why \texttt{PLAN} compares $R_{\rm Hill}$ to $\Delta x$.  While this work was in progress, \texttt{PLAN} was already used in the analyses of \citet{Abod2019}, and  \citet{Nesvorny2019}.

Our former clump-finding tool analyzed the surface density of solids in the ($x,y$) cells of the hydrodynamic grid.  This technique identifies prominent clumps and calculates the clump masses as the encapsulated column mass within their projected Hill radii.  Such a treatment is limited by the grid resolution and has difficulty detecting small planetesimals, especially when a massive clump is nearby.  Consequently, this method tends to overestimate the clump mass by including the mass of surrounding small planetesimals as well as other solids that are vertically far away.  \texttt{PLAN} overcomes those issues by diagnosing the particle data instead, as described above.

Fig. \ref{fig:CDF} shows that the \texttt{PLAN} results agree with our previous results at large masses, though the previous analyses slightly overestimated masses, as explained above.  The significant advantage of the \texttt{PLAN} analysis is that we identify gravitationally bound clumps at much lower masses than before, which improves our ability to statistically characterize the resulting mass distributions.

%%%%%%%%%%%%%%%%%%%%%%%%%%%%%%%%%%%%%%%%%%%%%%%%%%%%%%%%%%%%%%%%%%%%%%%%%%%%%%%%
\section{Statistical Modeling of the Mass Distribution}
\label{sec:fitting}

This section details our statistical methodology for analyzing the planetesimal mass distribution.  We present a maximum likelihood estimator (MLE) for estimating the parameters of a given model, and the uncertainty on those parameters in Section \ref{subsec:MLE}.  Section \ref{subsec:models} lists the models that we fit, which vary in complexity from 2 to 5 parameters.  Finally, Section \ref{subsec:select} describes our model selection criteria, and how they apply a penalty to more complex models.

\subsection{Maximum Likelihood Parameter Estimation}
\label{subsec:MLE}

We assume that the masses, $M$, of planetesimals (strictly, protoplanetesimal clumps in the simulations) are drawn from a probability density function (PDF), $\xi(M)$, parameterized by a vector $\bm{\theta}$, where $\xi(M; \bm{\theta})dM$ represents the probability that a given clump forms in the mass interval $M$ to $M+dM$.

In practice, it is easier to work with a logarithmic mass coordinate,  $x = \ln{(M/M_{\rm min})}$, referenced to the minimum mass of the distribution, $M_{\rm min}$.  With this transformation, the functional form of the PDF is different and written as
\begin{equation}
  p(x; \bm{\theta}) = \frac{1}{N_{\rm tot}} \fracd{N(x; \bm{\theta})}{x} = C(\bm{\theta}) g(x; \bm{\theta}),
\end{equation}
where the first equality simply relates the PDF to the total number of bodies $N_{\rm tot}$ and the number in a given logarithmic mass interval, $dN$.  In the second equality, the normalization factor $C(\bm{\theta})$ is introduced for later convenience (see also \citealt{Youdin2011b}).

Accordingly, the cumulative distribution function (CDF) is
\begin{equation}
  P_>(x; \bm{\theta}) = \int\limits_x^{+\infty} p(x'; \bm{\theta}) dx' = C(\bm{\theta}) \int\limits_x^{+\infty} g(x'; \bm{\theta}) dx',
\end{equation}
which denotes the expected fraction of clumps with masses larger than $M (= e^{x}M_{\rm min})$ in the distribution\footnote{Note the minus sign when relating the PDF to the CDF: $\displaystyle{p(x;\bm{\theta}) = -\fracd{}{x}P_>(x;\bm{\theta})}$.}.  The normalization of the CDF, $P_>(x; \bm{\theta})|_{x = 0} = 1$, gives
\begin{equation}
  C(\bm{\theta}) = \frac{1}{\int\limits_0^{+\infty} g(x; \bm{\theta}) dx},
\end{equation}
which requires that $g(x; \bm{\theta})$ does not diverge as $x\to+\infty$.

To fit a model to the data, i.e.\ the simulated mass distribution of planetesimals, we consider the \emph{likelihood} of the data given the model
\begin{equation}
  \mathcal{L}(\bm{x}|\bm{\theta}) \equiv \prod\limits_{i=1}^{N_{\rm tot}} p(x_i; \bm{\theta}) = \prod\limits_{i=1}^{N_{\rm tot}} C(\bm{\theta}) g(x_i; \bm{\theta}).
\end{equation}
It is usually easier to consider the log-likelihood
\begin{equation}
  \ln\mathcal{L}(\bm{x}|\bm{\theta}) = \sum\limits_{i=1}^{N_{\rm tot}} \ln p(x_i; \bm{\theta}) = N_{\rm tot} \ln{C(\bm{\theta})} + \sum\limits_{i=1}^{N_{\rm tot}} \ln g(x_i; \bm{\theta}).
\end{equation}

The maximum likelihood estimator (MLE) of $\bm{\theta}$ estimates the best-fit parameters, $\bm{\theta}_{\rm MLE}$, by maximizing the log-likelihood (within certain physical bounds if necessary).  For some simple log-likelihood functions, $\bm{\theta}_{\rm MLE}$ can be solved 
as the root(s) of
\begin{equation}\label{eq:ddL}
  \left\{\begin{aligned}
    \fracp{\ln\mathcal{L}(\bm{x}|\bm{\theta})}{\theta_j} &= 0, \\
    \fracpp{\ln\mathcal{L}(\bm{x}|\bm{\theta})}{\theta_j} &< 0,
  \end{aligned}\right.
\end{equation}
where $\theta_j$ means the $j$-th parameter in $\bm{\theta}$.  However, constraints on the allowed values of parameters sometimes confound traditional root-finding methods.

In this work, we apply numerical techniques to maximize the likelihood of the trial PDF (see Section \ref{subsec:models} for our choices).  In practice, we first use the python package \texttt{emcee} to explore the parameter space with a Markov-chain Monte Carlo (MCMC) approach \citep{Foreman-Mackey2013} to obtain an initial guess of parameters, $\bm{\theta}_{\rm MCMC}$.  We then use the \texttt{minimize} method\footnote{A complicated PDF may lead to a non-convex or non-smooth log-likelihood function, which is known to be difficult to minimize.  We always test different algorithms (e.g., ``Powell'', ``Newton-CG'', ``L-BFGS-B'', etc.) provided by \texttt{minimize} and run a set of optimizations with initial guesses selected in a mesh grid centered on $\bm{\theta}_{\rm MCMC}$.  We then take the solution leading to the lowest $-\ln\mathcal{L}$.} in the \texttt{scipy.optimize} package \citep{Numpy} to find the most likely $\bm{\theta}$ that minimizes $-\ln\mathcal{L}(\bm{x}|\bm{\theta})$.

To quantify the uncertainties of the best-fit parameters, $\bm{\theta}_{\rm MLE}$, we adopt the \textit{nonparametric bootstrap method} \citep{Efron1994, Burnham2002}.  By repeatedly taking a random sample of size $N_{\rm tot}$ \emph{with replacement} from the actual mass data, we first generate $N_{\rm bs}$ independent bootstrap samples.  They serve as a proxy for a set of $N_{\rm bs}$ independent real samples from the same mass distribution, because taking extra data (from additional simulations) is too costly.  The MLE is then employed to fit the model PDF to each bootstrap sample to obtain the best-fit parameters, $\bm{\theta}_{\mathrm{bs}, k}$ ($k=1, \cdots, N_{\rm bs}$).  Parameter uncertainties expected from real samples are estimated by calculating the distance between $\bm{\theta}_{\rm MLE}$ and the 84th and 16th percentiles of the distribution of $\bm{\theta}_{\mathrm{bs}, k}$, i.e. $\bm{\theta}_{\mathrm{bs}, k}^{84\%}$ and $\bm{\theta}_{\mathrm{bs}, k}^{16\%}$, as
\begin{equation}\label{eq:BMS_sigma}
  \begin{aligned}
    \Delta \bm{\theta}_{\rm bs}^+ &= \bm{\theta}_{\mathrm{bs}, k}^{84\%} - \bm{\theta}_{\rm MLE}, \\
    \Delta \bm{\theta}_{\rm bs}^- &= \bm{\theta}_{\rm MLE} - \bm{\theta}_{\mathrm{bs}, k}^{16\%}.
  \end{aligned}
\end{equation}
\citet{Efron1994} have shown that this bootstrap method works reasonably well if $N_{\rm bs}$ is large (e.g., $>$1000).  In this work, we fix $N_{\rm bs} = 10000$.  Appendix \ref{appsec:eg_fit} gives a model fitting example in detail.

Our maximum likelihood estimator is similar to what used in \citet{Simon2016, Simon2017, Abod2019} but we use bootstrapping to estimate parameter uncertainties.  A different method of parameter estimation is used by \citet{Johansen2015, Schafer2017}, etc., who applied curve fitting routines to the CDF.  While maximizing the likelihood functions seems more fundamental to us, we make no claim that our method is actually superior.  We conduct tests to recover the slope of a single power law distribution from randomly generated data using these two methods.  This test yields non-identical results, which confirms the methods are not equivalent, but offers no evidence to clearly favor either method.  A more rigorous investigation of this statistical issue could be warranted.

\subsection{Statistical Models}
\label{subsec:models}

We now describe the seven statistical models that are used to fit the mass distribution of planetesimals.  To focus on the shapes of these models, this section only gives their basic functional forms.  All the normalization coefficients and the full functional expressions are put in the Appendix \ref{appsec:model_coeff}.  For convenience, we define $K$ as the number of parameters in a model.

The first three models below are presented as CDFs.  We simply convert them to their corresponding PDFs to apply our maximum likelihood estimator.  However, it is natural to expect, if the planetesimal masses arise probabilistically, that a continuous PDF would be a more physical description.  The reason to consider the CDFs is that some appeared previously in the literature (the first two models) and one of our runs shows visual evidence for a kink in the CDF (the third model).  Because this kink gives a discontinuity in the PDF, it is arguably unphysical, but in this work we only examine statistical robustness, as no comprehensive physical theory for the distribution of masses exists.

\begin{enumerate}
  \item \textit{Simply Tapered Power Law} The concave downward profile of the CDFs of clump masses (see Fig. \ref{fig:CDF}) suggests a power law distribution with exponentially tapering \citep{Abod2019}
  \begin{equation}
    P_>(M; \bm{\theta})=c_1 M^{-\alpha}\ \exp\left(-\frac{M}{M_{\rm exp}}\right),
  \end{equation}
  where $M_{\rm exp}$ is the characteristic mass scale and $c_1$ is the renormalization coefficient (see Appendix \ref{appsec:model_coeff}, same for coefficients below).  This model has two free parameters ($K=2$) and $\bm{\theta}=(\alpha, M_{\rm exp})$, with $M_{\rm min} \leqslant M_{\rm exp} \leqslant M_{\rm max}$ but no constraints on $\alpha$.
  \item \textit{Variably Tapered Power Law} In addition to the first model, this model frees the tapering power by adding one more free parameter $\beta$ inside the exponential function \citep{Schafer2017}, 
  \begin{equation}
    P_>(M; \bm{\theta})=c_2 M^{-\alpha}\ \exp\left[-\left(\frac{M}{M_{\rm exp}}\right)^\beta\right],
  \end{equation}
  where $\bm{\theta}=(\alpha, \beta, M_{\rm exp})$ and $K=3$.  This model requires that at least one of $\alpha$ and $\beta$ is positive, and again $M_{\rm min} \leqslant M_{\rm exp} \leqslant M_{\rm max}$.
  \item \textit{Broken Cumulative Power Law} A broken cumulative power law distribution connects two power law segments with different slopes in the \emph{cumulative} distribution. It also manifests different behaviors at different mass ranges
  \begin{equation}
    P_>(M; \bm{\theta}) = \left\{\begin{aligned}
      & c_{31} M^{-\alpha_1}\ &M\leqslant M_{\rm br} \\
      & c_{32} M^{-\alpha_2}\ &M> M_{\rm br}
    \end{aligned}\right.,
  \end{equation}
  where $M_{\rm br}$ denotes the characteristic mass scale at which the slope breaks.  This model has three free parameters ($K=3$) and $\bm{\theta} = (\alpha_1, \alpha_2, M_{\rm br})$.  There is no constraints on $\alpha_1$, but $\alpha_2$ needs to be positive, and $M_{\rm min} \leqslant M_{\rm br} \leqslant M_{\rm max}$.
  \item \textit{Truncated Power Law} \citet{Schafer2017} also tested a truncated power law model
  \begin{equation}
    \xi(M; \bm{\theta}) = \left\{\begin{aligned}
      &c_4 M^{-\alpha-1}\ &M\leqslant M_{\rm tr} \\
      &0\ &M> M_{\rm tr}
    \end{aligned}\right.,
  \end{equation}
  where an upper bound $M_{\rm tr}$ truncates the PDF (and CDF).  In this model, $\bm{\theta} = (\alpha, M_{\rm tr})$, $K=2$, $\alpha > 0$, and $M_{\rm tr} \geqslant M_{\rm max}$.  Here the power law exponent in PDF becomes $-\alpha-1$ because the exponent in the corresponding CDF is $-\alpha$.  Furthermore, it is easy to show that the PDF monotonically increases with increasing $M_{\rm tr} \leq M_{\rm max}$, and hence $-\ln\mathcal{L}(\bm{x}|\bm{\theta})$ minimizes if and only if $M_{\rm tr} = M_{\rm max}$.
  \item \textit{Broken Power Law} Another compelling possibility is the broken power law distribution\footnote{Not to be confused with the broken cumulative power law.}.  The corresponding PDF consists of two different power law segments, leading to a smooth transition in the CDF near the breaking point
  \begin{equation}
    \xi(M; \bm{\theta}) = \left\{\begin{aligned}
      & c_{51} M^{-\alpha_1-1}\ &M\leqslant M_{\rm br} \\
      & c_{52} M^{-\alpha_2-1}\ &M> M_{\rm br}
    \end{aligned}\right. .
  \end{equation}
  This model has three free parameters ($K=3$), $\bm{\theta} = (\alpha_1, \alpha_2, M_{\rm br})$, where $\alpha_1$ has no limits, $\alpha_2 > 0$, and $M_{\rm min} \leqslant M_{\rm br} \leqslant M_{\rm max}$.  When $\alpha_2\to+\infty$, this model reverts to the Truncated Power Law model.
  \item \textit{Truncated Broken Power Law} This model complicates the last model by adding a truncation to the PDF
  \begin{equation}
    \xi(M; \bm{\theta}) = \left\{\begin{aligned}
      & c_{61} M^{-\alpha_1-1}\ &M\leqslant M_{\rm br} \\
      & c_{62} M^{-\alpha_2-1}\ &\text{otherwise} \\
      & 0\ &M> M_{\rm tr} 
    \end{aligned}\right. .
  \end{equation}
  This model has four free parameters ($K=4$) and $\bm{\theta} = (\alpha_1, \alpha_2, M_{\rm br}, M_{\rm tr})$, where $\alpha_1$ and $\alpha_2$ has no limits, $M_{\rm min} \leqslant M_{\rm br} \leqslant M_{\rm max} \leqslant M_{\rm tr}$.  Similar to the Truncated Power Law model, the PDF monotonically decreases with $M_{\rm tr}$ and $-\ln\mathcal{L}(\bm{x}|\bm{\theta})$ minimizes when $M_{\rm tr} = M_{\rm max}$.
  \item \textit{Three-segment Power Law} We take a step further to consider another broken power law distribution but with three segments in the PDF,
  \begin{equation}
    \xi(M; \bm{\theta}) = \left\{\begin{aligned}
      & c_{71} M^{-\alpha_1-1}\ &M\leqslant M_{\rm br1} \\
      & c_{72} M^{-\alpha_2-1}\ &\text{otherwise} \\
      & c_{73} M^{-\alpha_3-1}\ &M>M_{\rm br2}
    \end{aligned}\right. .
  \end{equation}
  This model has five free parameters ($K=5$) and $\bm{\theta} = (\alpha_1, \alpha_2, \alpha_3, M_{\rm br1}, M_{\rm br2})$.  Both $\alpha_1$ and $\alpha_2$ have no boundaries, but $\alpha_3 > 0$ and $M_{\rm min} \leqslant M_{\rm br1} \leqslant M_{\rm br2} \leqslant M_{\rm max}$.  When $\alpha_3\to+\infty$, this model reverts to the Truncated Broken Power Law model.
\end{enumerate}

We choose these seven statistical models as candidates since they have been previously used to fit the planetesimal mass function or are commonly applied to fit top-heavy mass distributions.  Other models are also certainly possible, but are beyond the scope of this paper.  Note that all the models above are transformed to a PDF function of $x$ (see Table \ref{tab:models}) to be used in our MLE.

\subsection{Model Selection Criteria}
\label{subsec:select}

Out next goal is to select the statistical models that best represent simulation data.  Models with more parameters (larger $K$) have the flexibility to provide closer fits to the data, i.e.\ higher likelihood values.  Often, a well-chosen function with fewer parameters can provide a better fit than a different function with more parameters.  The much larger concern is the opposite case, where a more complex model does not better represent reality, but merely overfits statistical fluctuations in the data.

For the problem of planetesimal formation by the streaming instability, this statistical concern is relevant.  The high mass tail of the planetesimal distribution is very significant, but with low numbers of high mass clumps in any simulation, the risk of statistical fluctuations impacting model fitting is potentially high.

To address these issues, we first review two of the most commonly-used model selection criteria and then introduce a selection criterion that we develop independently, motivated by the nonparametric bootstrap method.

\begin{deluxetable}{c|c}
  \tabletypesize{\normalsize}
  \tablecaption{Interpretation Guidelines for BIC \label{tab:BIC_guide}}
  \tablecolumns{2}
  \tablehead{
    \colhead{$\Delta_{\rm BIC}$} &
    \colhead{Evidence against the preferred model}
  }
  \startdata
  \hline\hline
  $0-2$  & Not worth more than a bare mention \\
  $2-6$  & Positive \\
  $6-10$ & Strong \\
  $>10$  & Very Strong
  \enddata
\end{deluxetable}

\begin{deluxetable}{c|c}
  \tabletypesize{\normalsize}
  \tablecaption{Interpretation Guidelines for AIC \label{tab:AIC_guide}}
  \tablecolumns{2}
  \tablehead{
    \colhead{$\Delta_{\rm AIC}$} &
    \colhead{Level of empirical support for a model}
  }
  \startdata
  \hline\hline
  $0-2$  & Substantial \\
  $2-4$  & Strong \\
  $4-7$  & Considerably less  \\
  $>10$  & Essentially none
  \enddata
\end{deluxetable}

\subsubsection{Information Criteria}
\label{subsubsec:aicbic}

The most commonly used model selection criteria are (i) the Bayesian Information Criterion (BIC) \citep{Kass1995}
\begin{equation}\label{eq:bic}
  \text{BIC} = K\ln(N_{\rm tot}) - 2 \ln\mathcal{L},
\end{equation}
and (ii) the Akaike Information Criterion (AIC) \citep{Akaike1974}
\begin{equation}\label{eq:aic}
  \text{AIC} = 2 K - 2 \ln\mathcal{L}.
\end{equation}
Both the BIC and the AIC involve the calculations of the log-likelihood $-2\ln{\mathcal{L}}$, which are affected by arbitrary constants and the sample size.  Thus, the individual BIC/AIC values are not significant and the relative differences between models
\begin{equation}
  \begin{aligned}
    \Delta_{\rm BIC} &= \text{BIC} - \text{BIC}_{\rm min}, \\
    \Delta_{\rm AIC} &= \text{AIC} - \text{AIC}_{\rm min}
  \end{aligned}
\end{equation}
are more important, where BIC$_{\rm min}$/AIC$_{\rm min}$ is the minimum of the BIC/AIC values of all the model candidates.  In this way, the preferred model naturally has $\Delta_{\rm BIC} = 0$ and other models have positive $\Delta_{\rm BIC}$'s (similar for AIC).  To interpret $\Delta_{\rm BIC}$ and $\Delta_{\rm AIC}$ quantitatively in model selection, we follow the conventional categorical guidelines in Tables \ref{tab:BIC_guide} and \ref{tab:AIC_guide}.

Formally, the value of $\Delta_{\rm BIC}$ represents the complexity-corrected likelihood ratio in a natural logarithmic scale, or the evidence provided by the data in favor of the preferred statistical model over another model \citep{Kass1995}.  The value of $\Delta_{\rm AIC}$ measures the Kullback–Leibler distance, or the information lost when a less preferred model is used to approximate the true distribution \citep{Burnham2002}.  For further discussions on the differences between the BIC and the AIC, we refer the reader to \citet{Burnham2002} and \citet{Burnham2004}.

These two criteria put different weights on the penalty on the number of parameters, $K$, which becomes quite significant for large $N_{\rm tot}$, and which can lead to different results in model selection.  It is difficult (for us) to determine which information criterion is more appropriate, or indeed if either is reliable.  More complex and computationally methods exist to assess the complexity penalty based not simply on the number of free parameters and/or data points, but the actual geometry of the model space \citep[e.g., Fisher Information Approximation,][]{Ly2017}.  However these methods were beyond the scope of this work.  Instead we describe an alternate model selection method below which we compare to the conventional AIC/BIC methods.

\subsubsection{Bootstrap Model Selection}
\label{subsubsec:bms}

Motivated by concerns about the applicability of standard model selection techniques (BIC and AIC, discussed above), we consider an alternative method where the complexity penalty is not given as a fixed, simple function of the number of parameters but instead is generated automatically by bootstrapping.

Inspired by the nonparametric bootstrap method for uncertainty estimation (described in Section \ref{subsec:MLE}), we again consider all the bootstrap samples a good proxy for mass distributions from $N_{\rm bs}$ independent simulations, which in reality are too computationally costly expensive to be conducted.  Through such a proxy, the median likelihood of all the bootstrap samples given the best-fit parameters can be used as a model selection criterion
\begin{equation}\label{eq:bms}
  \text{BMS} = -2\times \text{median}\left(\ln\mathcal{L}(\bm{x}_k|\bm{\theta}_{\rm MLE})\right).
\end{equation}
where BMS stands for Bootstrap Model Selection, $\bm{x}_k$ is the $k$-th bootstrap sample, and the factor of $2$ is chosen for similarity to AIC/BIC.  This empirical criterion represents to what extent the best-fit parameters can explain/describe other samples drawn from the same mass distribution.  Also, it naturally penalizes more complex models that tend to overfit data because they yield poorer fits to those bootstrap samples that deviate farther from the original data.  In the following work, we therefore also consider
\begin{equation}
  \Delta_{\rm BMS} = \text{BMS} - \text{BMS}_{\rm min},
\end{equation}
as one of our model selection metrics and follow the similar conventional categorical guidelines.  The comparison between BMS and the commonly-used BIC/AIC are discussed in the following sections.

%%%%%%%%%%%%%%%%%%%%%%%%%%%%%%%%%%%%%%%%%%%%%%%%%%%%%%%%%%%%%%%%%%%%%%%%%%%%%%%%

\begin{deluxetable*}{c|c|c|c|c|c|c}%[!htbp]
  \tablecaption{Model Fitting Results for Run I ($t_{\rm sg} = 7.5/\Omega$) \label{tab:fit_A}}
  %\tabletypesize{\small}
  \tablecolumns{7}
  \tablehead{
    \colhead{Models} &
    \colhead{Best-fit Parameters} &
    \colhead{Mass Scales [$M_G$]} &
    \colhead{$-\ln{\mathcal{L}}$} & %(\bm{x}|\bm{\theta}_{\rm MLE})}$}} &
    \colhead{$\Delta_{\rm BMS}$} &
    \colhead{$\Delta_{\rm BIC}$} &
    \colhead{$\Delta_{\rm AIC}$}
  }
  \startdata
  \hline\hline
  \begin{minipage}[c][1.2cm][c]{0.275\textwidth}
    \centering Simply Tapered Power Law \\ K=2, $\bm{\theta}=(\alpha, x_{\rm exp})$
  \end{minipage} &
  \begin{tabular}[c]{l}
    $\alpha = 0.208^{+0.011}_{-0.014}$ \\
    $x_{\rm exp} = 8.905^{+0.323}_{-0.464}$
  \end{tabular} &
  $M_{\rm exp} = 0.1385^{+0.0529}_{-0.0515}$ & 660.443 &
  63.7 &       53.1 &       60.4  \\
  \hline
  \begin{minipage}[c][1.7cm][c]{0.275\textwidth}
    \centering Variably Tapered Power Law \\ K=3, $\bm{\theta}=(\alpha, \beta, x_{\rm exp})$
  \end{minipage} &
  \begin{tabular}[c]{l}
    $\alpha = 0.036^{+0.041}_{-0.041}$ \\
    $\beta = 0.298^{+0.061}_{-0.040}$ \\
    $x_{\rm exp} = 4.734^{+1.022}_{-1.128}$
  \end{tabular} &
  $M_{\rm exp} = 0.0021^{+0.0038}_{-0.0014}$ & 633.695 &
  10.6 &        5.3 &        8.9  \\
  \hline
  \begin{minipage}[c][1.7cm][c]{0.275\textwidth}
    \centering Broken Cumulative Power Law \\ K=3, $\bm{\theta}=(\alpha_1, \alpha_2, x_{\rm br})$
  \end{minipage} &
  \begin{tabular}[c]{l}
    $\alpha_1 = 0.162^{+0.009}_{-0.019}$ \\
    $\alpha_2 = 0.589^{+0.042}_{-0.071}$ \\
    $x_{\rm br} = 4.550^{+0.003}_{-0.607}$
  \end{tabular} &
  $M_{\rm br} = 0.0018^{+0.0000}_{-0.0008}$ & 631.683 &
  10.1 &        1.2 &        4.9  \\
  \hline\hline
  \begin{minipage}[c][1.2cm][c]{0.275\textwidth}
    \centering Truncated Power Law \\ K=2, $\bm{\theta}=(\alpha, x_{\rm tr})$
  \end{minipage} &
  \begin{tabular}[c]{l}
    $\alpha = 0.140^{+0.012}_{-0.029}$ \\
    $x_{\rm tr} = 10.880^{+0.000}_{-0.000}$
  \end{tabular} &
  $M_{\rm tr} = 0.9981^{+0.0000}_{-0.0000}$ & 651.846 &
  47.5 &       35.9 &       43.2  \\
  \hline
  \begin{minipage}[c][1.7cm][c]{0.275\textwidth}
    \centering Broken Power Law \\ K=3, $\bm{\theta}=(\alpha_1, \alpha_2, x_{\rm br})$
  \end{minipage} &
  \begin{tabular}[c]{l}
    $\mathbf{\alpha_1 = -0.079^{+0.043}_{-0.049}}$ \\
    $\alpha_2 = 0.628^{+0.078}_{-0.055}$ \\
    $x_{\rm br} = 5.620^{+0.329}_{-0.285}$
  \end{tabular} &
  $M_{\rm br} = 0.0052^{+0.0020}_{-0.0013}$ & 631.946 &
   7.2 &        1.8 &        5.4  \\
  \hline
  \begin{minipage}[c][2.7cm][c]{0.275\textwidth}
    \centering Truncated Broken Power Law \\ K=4, $\bm{\theta}=(\alpha_1, \alpha_2, x_{\rm br}, x_{\rm tr})$
  \end{minipage} &
  \begin{tabular}[c]{l}
    $\mathbf{\alpha_1 = -0.102^{+0.058}_{-0.043}}$ \\
    $\alpha_2 = 0.468^{+0.082}_{-0.070}$ \\
    $x_{\rm br} = 5.066^{+0.519}_{-0.144}$ \\
    $x_{\rm tr} = 10.880^{+0.000}_{-0.000}$
  \end{tabular} &
  \begin{tabular}[c]{l}
    $M_{\rm br} = 0.0030^{+0.0020}_{-0.0004}$\\
    $M_{\rm tr} = 0.9981^{+0.0000}_{-0.0000}$
  \end{tabular} & 628.241 &
  \textbf{0.0} &        \textbf{0.0} &        \textbf{0.0} \\
  \hline
  \begin{minipage}[c][2.7cm][c]{0.275\textwidth}
    \centering Three-segment Power Law \\ K=5, $\bm{\theta}=(\alpha_1, \alpha_2, \alpha_3, x_{\rm br1}, x_{\rm br2})$
  \end{minipage} &
  \begin{tabular}[c]{l}
    $\mathbf{\alpha_1 = -0.102^{+0.057}_{-0.043}}$ \\
    $\alpha_2 = 0.468^{+0.081}_{-0.071}$ \\
    $\alpha_3 = 5.78\se5^{+6.07\se7}_{-3.9\se4}\tablenotemark{*}$ \\
    $x_{\rm br1} = 5.066^{+0.511}_{-0.142}$ \\
    $x_{\rm br2} = 10.880^{+0.000}_{-0.632}$
  \end{tabular} &
  \begin{tabular}[c]{l}
    $M_{\rm br1} = 0.0030^{+0.0020}_{-0.0004}$\\
    $M_{\rm br2} = 0.9981^{+0.0000}_{-0.4674}$
  \end{tabular} & 628.241 &
  \textbf{0.0} &        5.6 &        2.0 \\
  \enddata
  \tablenotetext{*}{The best-fit third slope, $\alpha_3$, of the Three-segment Power Law model is extremely large because this model essentially reverts to the Truncated Broken Power Law model (see also Section \ref{subsec:models} and Fig. \ref{fig:fit_AC}).}
\end{deluxetable*}

\begin{deluxetable*}{c|c|c|c|c|c|c}%[!htbp]
  \tablecaption{Model Fitting Results for Run II ($t_{\rm sg} = 7.6/\Omega$) \label{tab:fit_C}}
  %\tabletypesize{\scriptsize}
  \tablecolumns{7}
  \tablehead{
    \colhead{Models} &
    \colhead{Best-fit Parameters} & 
    \colhead{Mass Scales [$M_G$]} &
    \colhead{$-\ln{\mathcal{L}}$} & %(\bm{x}|\bm{\theta}_{\rm MLE})}$}} &
    \colhead{$\Delta_{\rm BMS}$} &
    \colhead{$\Delta_{\rm BIC}$} &
    \colhead{$\Delta_{\rm AIC}$}
  }
  \startdata
  \hline\hline
  \begin{minipage}[c][1.2cm][c]{0.275\textwidth}
    \centering Simply Tapered Power Law \\ K=2, $\bm{\theta}=(\alpha, x_{\rm exp})$
  \end{minipage} &
  \begin{tabular}[c]{l}
    $\alpha = 0.388^{+0.030}_{-0.039}$ \\
    $x_{\rm exp} = 6.397^{+0.473}_{-0.721}$
  \end{tabular} &
  $M_{\rm exp} = 11.2531^{+6.8113}_{-5.7787}$ & 311.520 &
  16.1 &       10.2 &       13.3  \\
  \hline
  \begin{minipage}[c][1.7cm][c]{0.275\textwidth}
    \centering Variably Tapered Power Law \\ K=3, $\bm{\theta}=(\alpha, \beta, x_{\rm exp})$
  \end{minipage} &
  \begin{tabular}[c]{l}
    $\alpha = 0.304^{+0.063}_{-0.060}$ \\
    $\beta = 0.527^{+0.233}_{-0.088}$ \\
    $x_{\rm exp} = 5.178^{+0.743}_{-0.777}$
  \end{tabular} &
  $M_{\rm exp} = 3.3255^{+3.6627}_{-1.7965}$ & 309.274 &
  11.6 &       10.8 &       10.8  \\
  \hline
  \begin{minipage}[c][1.7cm][c]{0.275\textwidth}
    \centering Broken Cumulative Power Law \\ K=3, $\bm{\theta}=(\alpha_1, \alpha_2, x_{\rm br})$
  \end{minipage} &
  \begin{tabular}[c]{l}
    $\alpha_1 = 0.348^{+0.035}_{-0.033}$ \\
    $\alpha_2 = 0.866^{+0.143}_{-0.080}$ \\
    $x_{\rm br} = 3.226^{+0.035}_{-0.065}$
  \end{tabular} &
  $M_{\rm br} = 0.4719^{+0.0169}_{-0.0298}$ & 303.864 &
   1.4 &        \textbf{0.0} &        \textbf{0.0}  \\
  \hline\hline
  \begin{minipage}[c][1.2cm][c]{0.275\textwidth}
    \centering Truncated Power Law \\ K=2, $\bm{\theta}=(\alpha, x_{\rm tr})$
  \end{minipage} &
  \begin{tabular}[c]{l}
    $\alpha = 0.360^{+0.029}_{-0.051}$ \\
    $x_{\rm tr} = 7.891^{+0.000}_{-0.000}$
  \end{tabular} &
  $M_{\rm tr} = 50.126^{+0.000}_{-0.000}$ & 310.769 &
  14.9 &        8.7 &       11.8  \\
  \hline
  \begin{minipage}[c][1.7cm][c]{0.275\textwidth}
    \centering Broken Power Law \\ K=3, $\bm{\theta}=(\alpha_1, \alpha_2, x_{\rm br})$
  \end{minipage} &
  \begin{tabular}[c]{l}
    $\alpha_1 = 0.240^{+0.064}_{-0.068}$ \\
    $\alpha_2 = 0.895^{+0.265}_{-0.103}$ \\
    $x_{\rm br} = 4.431^{+0.322}_{-0.180}$
  \end{tabular} &
  $M_{\rm br} = 1.5754^{+0.5992}_{-0.2594}$ & 309.058 &
  11.0 &       10.4 &       10.4  \\
  \hline
  \begin{minipage}[c][2.7cm][c]{0.275\textwidth}
    \centering Truncated Broken Power Law \\ K=4, $\bm{\theta}=(\alpha_1, \alpha_2, x_{\rm br}, x_{\rm tr})$
  \end{minipage} &
  \begin{tabular}[c]{l}
    $\alpha_1 = 0.249^{+0.065}_{-0.077}$ \\
    $\alpha_2 = 0.729^{+0.152}_{-0.200}$ \\
    $x_{\rm br} = 4.330^{+0.261}_{-0.691}$ \\
    $x_{\rm tr} = 7.891^{+0.000}_{-0.000}$
  \end{tabular} &
  \begin{tabular}[c]{l}
    $M_{\rm br} = 1.4241^{+0.4254}_{-0.7103}$\\
    $M_{\rm tr} = 50.1262^{+0.0000}_{-0.0000}$
  \end{tabular} & 307.718 &
   8.4 &       12.9 &        9.7 \\
  \hline
  \begin{minipage}[c][2.7cm][c]{0.275\textwidth}
    \centering Three-segment Power Law \\ K=5, $\bm{\theta}=(\alpha_1, \alpha_2, \alpha_3, x_{\rm br1}, x_{\rm br2})$
  \end{minipage} &
  \begin{tabular}[c]{l}
    $\alpha_1 = 0.771^{+0.291}_{-0.159}$ \\
    $\alpha_2 = -0.362^{+0.181}_{-0.303}$ \\
    $\alpha_3 = 0.833^{+0.164}_{-0.075}$ \\
    $x_{\rm br1} = 1.827^{+0.270}_{-0.314}$ \\
    $x_{\rm br2} = 3.501^{+0.198}_{-0.079}$
  \end{tabular} &
  \begin{tabular}[c]{l}
    $M_{\rm br1} = 0.1166^{+0.0362}_{-0.0314}$\\
    $M_{\rm br2} = 0.6216^{+0.1359}_{-0.0474}$
  \end{tabular} & 303.711 &
  \textbf{0.0} &       10.0 &        3.7
  \enddata
\end{deluxetable*}

\begin{figure*}
  \centering
  \includegraphics[width=\linewidth]{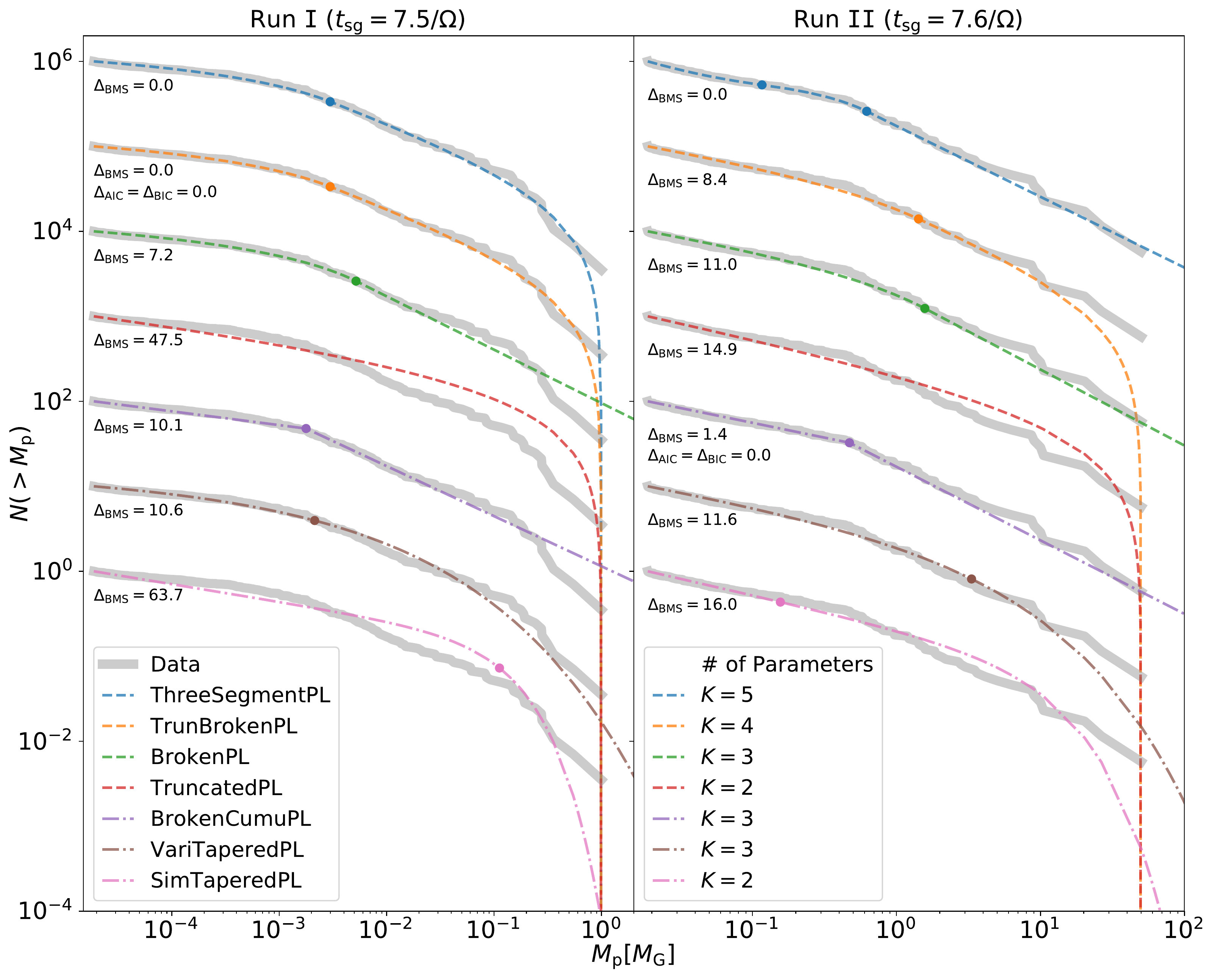}
  \caption{Fitting results to the simulated mass distribution in Run I (\textit{left}) and Run II (\textit{right}) at a similar $t_{\rm sg}$.  The resulting model CDFs (dashed curves) are overplotted on the simulation data (grey-shaded curves).  Each model is offset by $10$ from bottom to top for better visual comparison (``PL'' stands for ``Power Law'').  The dot(s) on each curve denote(s) the model-specific characteristic mass scale(s) as defined in Table \ref{tab:models} and listed in Tables \ref{tab:fit_A} and \ref{tab:fit_C} (no dot at the truncation mass ($M_{\rm tr}$) because the CDF decreases to $0$).  We emphasize that these fitting results are obtained by the maximum likelihood estimator described in Section \ref{subsec:MLE}.  The values of $\Delta_{\rm BMS}$ are annotated for reference. Models annotated with $\Delta_{\rm BMS/AIC/BIC}=0.0$ are preferred models by BMS/AIC/BIC. \label{fig:fit_AC}}
\end{figure*}

\begin{figure*}
  \centering
  \includegraphics[width=\linewidth]{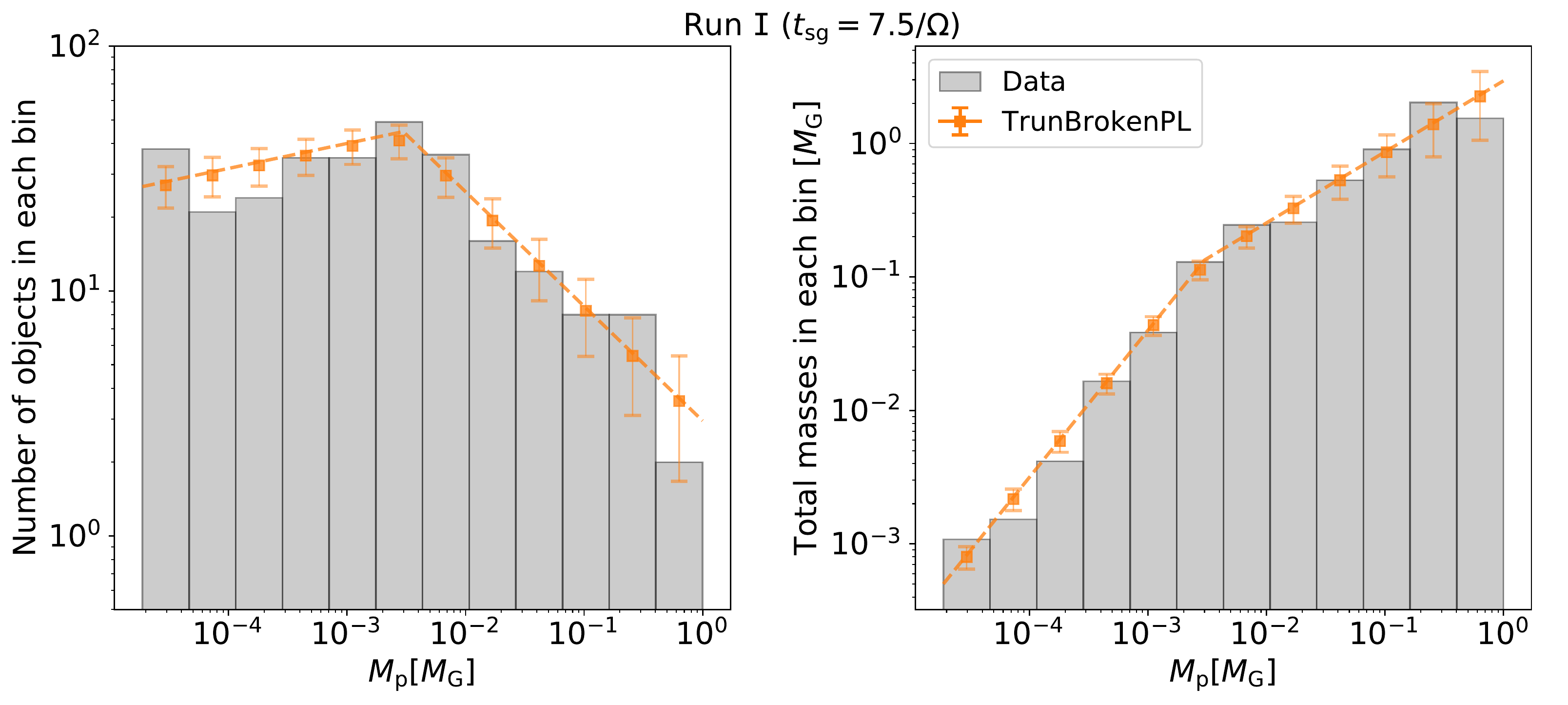} \\
  \includegraphics[width=\linewidth]{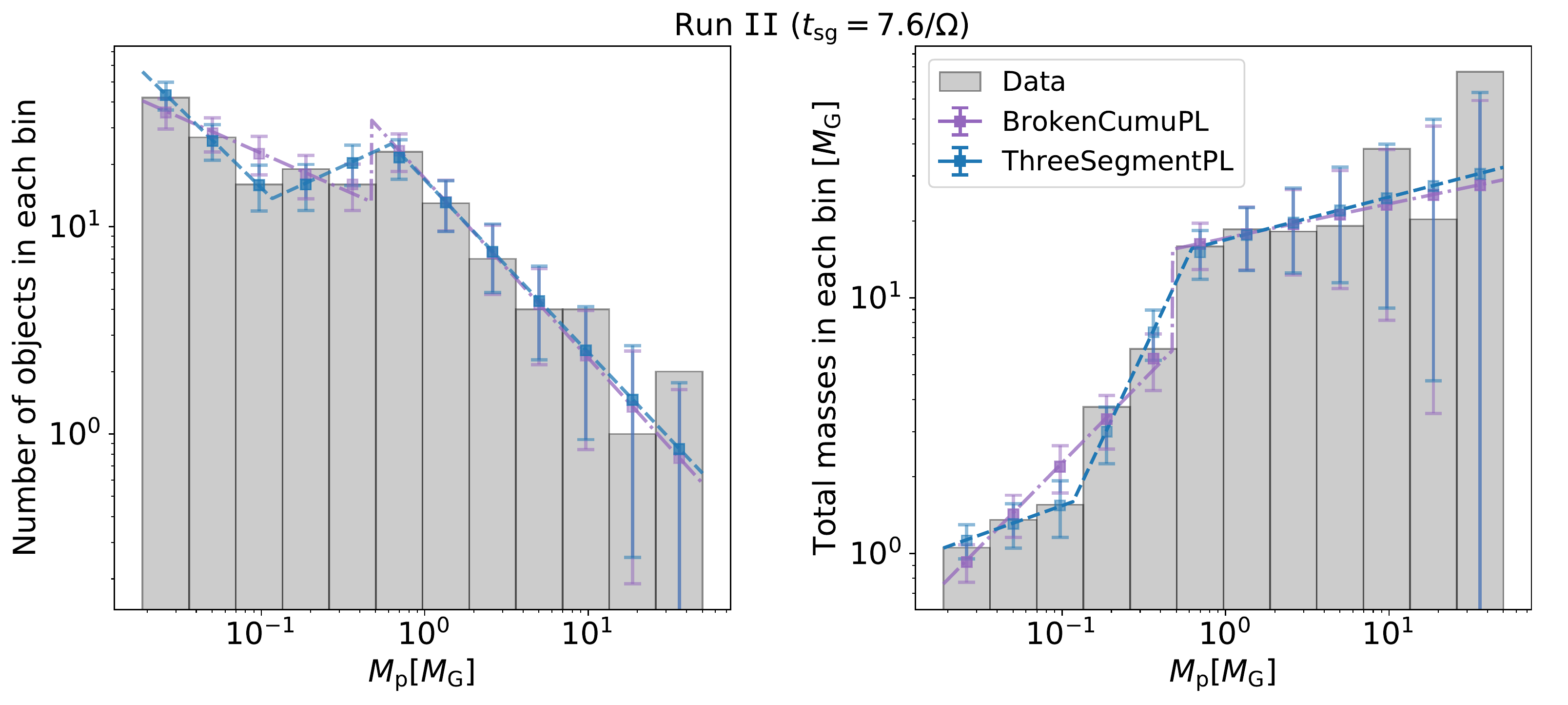}
  \caption{The number of clumps formed per unit logarithmic mass interval (left) and the total mass of clumps in each interval (right) for Run I (upper) and Run II (lower).  The PDFs of the preferred model(s) are overplotted for comparison, with error bars indicating the expected Poisson fluctuations.  More specifically, the dashed and dashed-dotted lines represent the analytical curves of the PDFs of the preferred model(s).  The points with error bars represent the average values of the PDFs over each mass intervals.  Thus, the analytical curves do not necessarily go through the center of each point. \label{fig:barplot}}
\end{figure*}

%%%%%%%%%%%%%%%%%%%%%%%%%%%%%%%%%%%%%%%%%%%%%%%%%%%%%%%%%%%%%%%%%%%%%%%%%%%%%%%%
\section{The Initial Planetesimal Mass Distribution}
\label{sec:results}

In this section, we analyze and compare our two high resolution simulations (Run I and II, see Table \ref{tab:setup}) that have different physical parameters $(\uptau_{\rm s}, Z)$.  We fit seven statistical models (described in Section \ref{subsec:models} and Appendix \ref{appsec:model_coeff}) to the simulated mass distribution of planetesimals identified by \texttt{PLAN} (described in Section \ref{subsec:plan}).  In this section, we first present the fitting results and then the preferred models by model selection criteria.  Section \ref{subsec:turnover} describes an interesting turnover in the fitted mass distribution of Run I.  In the end, Section \ref{subsec:comp2previous} compares our results to recent studies.

\subsection{Maximum Likelihood Fitting Results}
\label{subsec:best-fits}

Tables \ref{tab:fit_A} and \ref{tab:fit_C} list the best-fit likelihood and parameters for all the statistical models for Run I and II, respectively.  The mass distribution used in fitting is extracted at $t_{\rm sg} = 7.5/\Omega_0$ for Run I and at $t_{\rm sg} = 7.6/\Omega_0$\footnote{Run I is the same simulation snapshot presented in Fig. 1, panel 2 in \citet{Simon2017}, but re-analyzed here by \texttt{PLAN}.} for Run II.  By this time hundreds of planetesimals have formed, and the time evolution of the mass distribution has slowed.  As shown in the two tables, for each model, the best-fit parameters for the two simulations are statistically different and are outside the uncertainty ranges of each other, indicating two different mass distributions are produced.

Fig. \ref{fig:fit_AC} visualizes all the resulting model CDFs for both of our simulations.  They produce a broad mass distribution spanning more than three orders of magnitude in mass.  However, Run I and II cover  different mass regimes due to the different choices of physical parameters and hence the disk conditions, which may also require differently shaped distribution functions to describe.  We emphasize that a physical understanding of these differences is elusive, hence our focus on statistics.

For Run I, we find the best-fit first-segment power law slopes ($-\alpha_1$) for the last three models in Table \ref{tab:fit_A} become positive, which is of great interest for understanding which planetesimal masses dominate in number counts, and will be discussed in Section \ref{subsec:turnover}.  In addition, the third-segment power law slope ($-\alpha_3$) for the Three-segment Power Law model is extremely steep.  As mentioned in Section \ref{subsec:models}, when $\alpha_3$ approaches $+\infty$, this model reverts to the Truncated Broken Power Law, with the other four parameters being identical between these two models. 

In this paper, we do not further consider the mass distributions in other snapshots and the possible variations with time, which belongs to future work.  \citet{Simon2017} used a single power law model to fit the mass distribution and found that the power law index remains relatively constant in time after an initially rapid collapse.  \citet{Schafer2017} used the Variably Tapered Power Law model and also found that the power law index, characteristic mass scale, and tapering exponent all remain approximately constant in time for five orbital periods.  Based on these results, we do not expect the mass distribution in our simulations evolve rapidly, i. e.\ on dynamical timescales, after the snapshots.

\subsection{Model Selection}
\label{subsec:model_sel}

Our model selection analyses are based on the information criteria described in Section \ref{subsec:model_sel}, i.e. $\Delta_{\rm BMS}$, $\Delta_{\rm BIC}$, and $\Delta_{\rm AIC}$ values, presented in Tables \ref{tab:fit_A} and \ref{tab:fit_C}.  We find that more complex models ($K\geqslant3$) are generally much preferred than the other two simpler models ($K=2$), regardless of the model selection criteria.  In other words, the Simply Tapered Power Law and the Truncated Power Law models are consistently less favored.

In the case of Run I, all the model selection criteria reach a consensus on choosing the $K=4$ Truncated Broken Power Law as the preferred model.  However, there is disagreement on the strength of preference, as we now explain.  In this specific case, the $K=5$ Three-segment Power Law model reverts to the preferred $K=4$ model, 
because the high mass power law is very steep (effectively a truncation) with the other four parameters identical.  Since the BMS does not count parameters, it does not distinguish between these identically shaped distributions.  However the AIC and BIC both apply a penalty for the 5th parameter, which is much more significant for the BIC.  While all  model selection metrics prefer the $K=4$ Truncated Broken Power Law, the BIC method (only) finds that the evidence against a pair of $K=3$ models -- the Broken Cumulative Power Law model ($\Delta_{\rm BMS} = 1.2$) and the Broken Power Law model ($\Delta_{\rm BMS} = 1.8$) -- is not significant. 

In the Run II case, both the BIC and the AIC designate the Broken Cumulative Power Law model as the preferred model, which only has a moderate complexity level among all the model candidates.  However, the BMS prefers the Three-segment Power Law models slightly more but also substantially supports the Broken Cumulative Power Law model.  The overall preference for a broken CDF model may not be surprising given that the mass distribution shows a kink in the CDF (see Fig. \ref{fig:fit_AC}), but is less physical intuitive since the planetesimal masses are expected to arise probabilistically and broken PDFs have been observed in the size-frequency distributions of asteroids and Kuiper Belt Objects (KBOs) \citep[e.g.][]{Morbidelli2009, Fraser2010, Shankman2013, Singer2019}.  More work is needed to further understand whether our case is special or not. 

Fig. \ref{fig:fit_AC} allows a visual inspection of our model fitting and selection results.  Not surprisingly, more complex models generally better fit the mass distributions.  The effect of the logarithmic number axis is worth emphasizing.  A small deviation from the (logarithmically plotted) CDF at the low mass end is more statistically significant than a larger deviation at the high mass end, where numbers, especially cumulative numbers, are much higher.  The advantage of CDF plots is that no binning choices are required.  While no data features are lost to binning, a disadvantage is the difficulty in interpreting slopes at the low mass end (where the CDF is near unity).

Fig. \ref{fig:barplot} provides an alternate view of binned planetesimal numbers (masses) which are compared to the PDFs of the best fit models.  The PDF of the Broken Cumulative Power Law model -- fit to Run II -- has a discontinuous value at the CDF bread.  As noted earlier, a physical explanation for such break does not exist.  Overall, the selected models are excellent fits to the binned data, especially when accounting for the error bars, which represent the Poisson noise on the expected number (and mass) of bodies per bin.  For Run II, the two preferred models seem to represent the mass distribution equally well.

\subsection{A Turnover in the Mass Distribution}
\label{subsec:turnover}

In SI simulations to date, the low mass end of the PDF is described by a power law with $\alpha > 0$, where $dN/d\ln(M) \propto M^{-\alpha}$ (most of these works, described below, use $p \equiv \alpha + 1$).  Such a slope cannot extend to arbitrarily low mass, because the total mass of planetesimals would diverge.  Sufficiently high resolution simulations should solve this problem, by revealing a low mass tail with $\alpha < 0$.   Such a measurement would determine the mass of planetesimals formed by SI that initially dominate by number.  We present here the first evidence of such a turnover.

In Fig. \ref{fig:barplot}, the mass frequency distribution of Run I  presents a turnover below $\sim 0.003M_{\rm G}$ (roughly $100$ km-sized objects for the disk model in Section \ref{subsec:setup}).  The number of clumps drops with decreasing mass, except for the lowest mass bin (more on this below).  

Our preferred statistical model confirms this visual evidence.
The $K=4$ Truncated Broken Power Law model has a positive low mass slope of $-\alpha_1=0.102$ (as does the in practice identical $K=5$  Three-segment Power Law model).  Our bootstrap error estimates (see Table \ref{tab:fit_A}) confirm the significance of the positive slope.  Moreover, the simpler $K=3$ Broken Power Law model (while not the most preferred model) also has a positive low mass slope ($-\alpha_1=0.079$), which is flatter, but agrees within statistical uncertainties.

The evidence for a mass turnover (i.e. a positive slope at low masses) in Run I is fairly compelling.  However, Run I also has an increase in the number of bodies in the lowest mass bin in Fig. \ref{fig:barplot}.  It is unclear if this increase has a physical origin, though we suspect that insufficient resolution on the smallest scales is an issue.  We note that the lower resolution Run II shows an uptick in the planetesimal numbers in the two lowest mass bins.

Future studies with higher resolution are required to better resolve the low mass planetesimals and better characterize the inevitable turnover in the gravitational collapse mass function at low masses. 

By contrast, for Run II resolution is not sufficient for evidence of a low mass turnover.  However, one of the preferred statistical model (the $K=5$ Three-segment Power Law model) reveals a positive slope at intermediate masses ($-\alpha_2 = 0.362$).  Whether this slope extends to the low-mass end and whether the uptick of the two lowest mass bins is numerical again require higher resolution studies.  Moreover, the binned mass distribution of Run II shows a much flatter slope at the high mass regime than that of Run I, which is also shown by the CDFs in Fig. \ref{fig:fit_AC}.  This comparison further demonstrates that the different physical conditions of our two simulations produce different mass distributions. 

\subsection{Comparison with Previous Studies}
\label{subsec:comp2previous}

In this section, we compare our fitting results to two recent works on planetesimal mass distribution that considered models beyond a single power law.

\citet{Schafer2017} ran a suite of simulations with different resolutions and box sizes, fixing the physical parameters, $(\uptau_{\rm s}, Z, \Pi, \tilde{G}) = (0.314, 0.02, 0.05, 0.318)$, i.e.\ similar to our Run II with weaker self-gravity.  They considered a two-parameter Truncated Power Law model and a three-parameter Variably Tapered Power Law model and found the latter provides a better fit.  Our analysis did not favor either of these models, especially for the similar run II.  For the tapering exponent (of the Variably Tapered Power Law), they fit $\beta \simeq 0.3 $ -- $ 0.4$, similar to our results ($0.298$ and $0.527$ for Run I and II, respectively).  \citet{Schafer2017} explained that with limited resolution ($\Delta x = H/640$ at best), their simulations did not produce enough planetesimals to constrain the shape of the CDF in the power law part, and thus the values of $\alpha$ and $M_{\rm exp}$.  Their work used a different code, PENCIL, and also used a sink-particle algorithm to handle bound clumps.  These differences are a useful check on numerical robustness.

\citet{Abod2019} used high-resolution ($\Delta x = H/2560$) simulations with $(\uptau_{\rm s}, Z, \tilde{G}) = (0.05, 0.1, 0.02)$ -- i.e.\ smaller solids in a lower mass disk than our runs -- to study how the initial mass distribution of planetesimals depends on the pressure gradient, $\Pi$, with values from $0$ to $0.1$.  
They found that the planetesimal mass function depends at most weakly on $\Pi$.  \citet{Abod2019} used a two-parameter Simply Tapered Power Law model to fit the simulation data.  Our analysis did not prefer this model, though it does have an advantage of simplicity.   They fit the power law exponent $\alpha\approx 0.3$ and the characteristic mass scale of $\sim 0.3 M_{\rm G}$ when $\Pi = 0.05$.    Our results give similar power law slopes ($\alpha=0.208$ for Run I and $0.388$ for Run II) and, for Run I a similar characteristic mass scale ($M_{\rm exp}=0.1385M_{\rm G}$).  Our Run II fit,  $M_{\rm exp}(=11.2531M_{\rm G})$, differs by a factor of $\sim 81$, for reasons that are not yet clear. 

Since there are always more than one difference in the physical conditions and only limited model candidates are considered, these comparisons are somewhat inconclusive.  Though costly, more parameter studies are needed to understand how the initial planetesimal mass function varies with each of the four physical parameters $(\uptau_{\rm s}, Z, \Pi, \tilde{G})$ and eventually how these parameter dependencies couple.

%%%%%%%%%%%%%%%%%%%%%%%%%%%%%%%%%%%%%%%%%%%%%%%%%%%%%%%%%%%%%%%%%%%%%%%%%%%%%%%
\section{Discussions And Conclusions}
\label{sec:final}

We investigate the mass distribution of planetesimals formed in high-resolution SI simulations.  This mass distribution is of great astrophysical interest since it provides insights for the observations of small bodies in the Solar System  (e.g., Cold Classical Kuiper Belt Objects, \citealp[etc.]{Nesvorny2019}) as well as for the modeling of subsequent protoplanet formation \citep[e.g.][]{Liu2019}.

In this work, we conduct SI simulations including particle self-gravity with the highest resolution to date, which produce broad mass distributions of planetesimals.  While such distributions are top-heavy in mass for all numerical resolution choices, higher resolution simulations probe the lower-mass tail that dominates planetesimal numbers.  We also develop and apply a new clump-finding tool, \texttt{PLAN} (described in Section \ref{subsec:plan}), to accurately identify self-bound clumps in our simulations and extract their mass distributions.  \texttt{PLAN} was used in previous work \citep{Abod2019, Nesvorny2019}, but the details of the method are presented here. 

We fit the mass distribution to statistical models with different parameterizations (described in Section \ref{subsec:models}.  To determine and select the preferred model from simulation data is a difficult statistical art, especially when different model candidates have different numbers of parameters.  Thus, this work considers a variety of model selection criteria: the commonly-used BIC and AIC, as well as a bootstrap model selection method that we call BMS.

Based on our analyses, we find that
\begin{itemize}
  \item Run I is best described by a four-parameter model, the Truncated Broken Power Law.
  \item For Run II (with smaller solids and a lower solid abundance than Run I) the preferred model depends on the model selection criterion used.  The AIC and BIC prefer a three-parameter Broken Cumulative Power Law.  The BMS selects a five-parameter model, the Three-Segment Power Law.
\end{itemize}

The interpretations and conclusions are drawn as follows:
\begin{enumerate}
  \item The initial mass distribution of planetesimals formed by the streaming instability is shown to be numerically robust for the high mass regimes, and is most probably also robust at lower masses.  Simulations with different numerical resolution (Run I and an equivalent run with lower resolution) show a similar mass distribution at the high mass end.  Higher resolution gives a correction at intermediate masses and an extension to lower masses.
  \item For different physical conditions, the initial mass distribution is \textit{not} universal.  While all cases produce top-heavy mass distributions with similar overall shapes, simulations with different physical parameters produce statistically different mass distributions.  Fitting the same model to different runs often yields different best-fit parameters, e.g. power law slopes and characteristic mass scales.  Moreover, the preferred models for different runs have different functional forms.  More work and more high-resolution simulations are needed to better understand the initial mass distribution.
  \item Our preferred models were not previously considered in the literature.   We analyze the models that were used in previous studies, and find alternate models which rank higher by all model selection criteria.  We make no claim to have found the optimal model, which we may not have considered and which may change as simulation data improves.
  \item We find evidence for a turnover in the mass frequency distribution at the low mass end.  This evidence is most prominent for Run I, where the PDF of the logarithmic masses transitions to a positive slope below $M \sim0.003M_{\rm G}$ at roughly 2-$\sigma$ significance in the estimated slope.  There is also statistical evidence for a turnover in the case of Run II, but only at intermediate masses.  To better characterize the turnover of initial planetesimal mass distributions, higher resolution simulations are required.
  \item The most complex model is not always selected as the preferred model.  This result emphasizes the importance of applying complexity penalties for model selection.
  \item Different model selection criteria disagree on both the absolute and relative rankings of different models.  It is often difficult to rigorously justify a single model selection criteria for a given (astrophysical) application.  Absent this justification, we generally recommend that multiple selection criteria be considered to increase confidence in model selection analyses. 
\end{enumerate}

\citet{Nesvorny2019} recently find that the clumps formed by the SI possess excess angular momenta and are likely to form binaries or multiple systems.  In that case, the mass distributions from our simulations may describe the mass function of binaries/systems, not individual objects.  This finding introduces corrections to the overall mass distribution and also some uncertainties to our modeling, which are beyond the scope of this work.  However, those corrections and uncertainties would be minor if all clumps eventually form equal-size binaries as proposed in \citet{Nesvorny2010}.

%%%%%%%%%%%%%%%%%%%%%%%%%%%%%%%%%%%%%%%%%%%%%%%%%%%%%%%%%%%%%%%%%%%%%%%%%%%%%%%%
\section*{Acknowledgements}

We thank Kaitlin Kratter, Paola Pinilla, Philip Pinto, and Peter Behroozi for useful discussions.  RL acknowledges support from NASA headquarters under the NASA Earth and Space Science Fellowship Program grant NNX16AP53H. ANY acknowledges partial support from NASA Astrophysics Theory Grant NNX17AK59G and from the NSF through grant AST-1616929.

\software{ATHENA \citep{Stone2008, Bai2010}, 
          Matplotlib \citep{Matplotlib}, 
          Numpy \& Scipy \citep{Numpy},
          emcee v3.0rc2 \citep{Foreman-Mackey2013},
          PLAN \citep{PLAN},
          Pyridoxine \citep{Pyridoxine}.}

%%%%%%%%%%%%%%%%%%%%%%%%%%%%%%%%%%%%%%%%%%%%%%%%%%

%%%%%%%%%%%%%%%%% APPENDICES %%%%%%%%%%%%%%%%%%%%%

\appendix
\section{A Model Fitting Example}
\label{appsec:eg_fit}

\begin{figure}
  \centering
  \includegraphics[width=\linewidth]{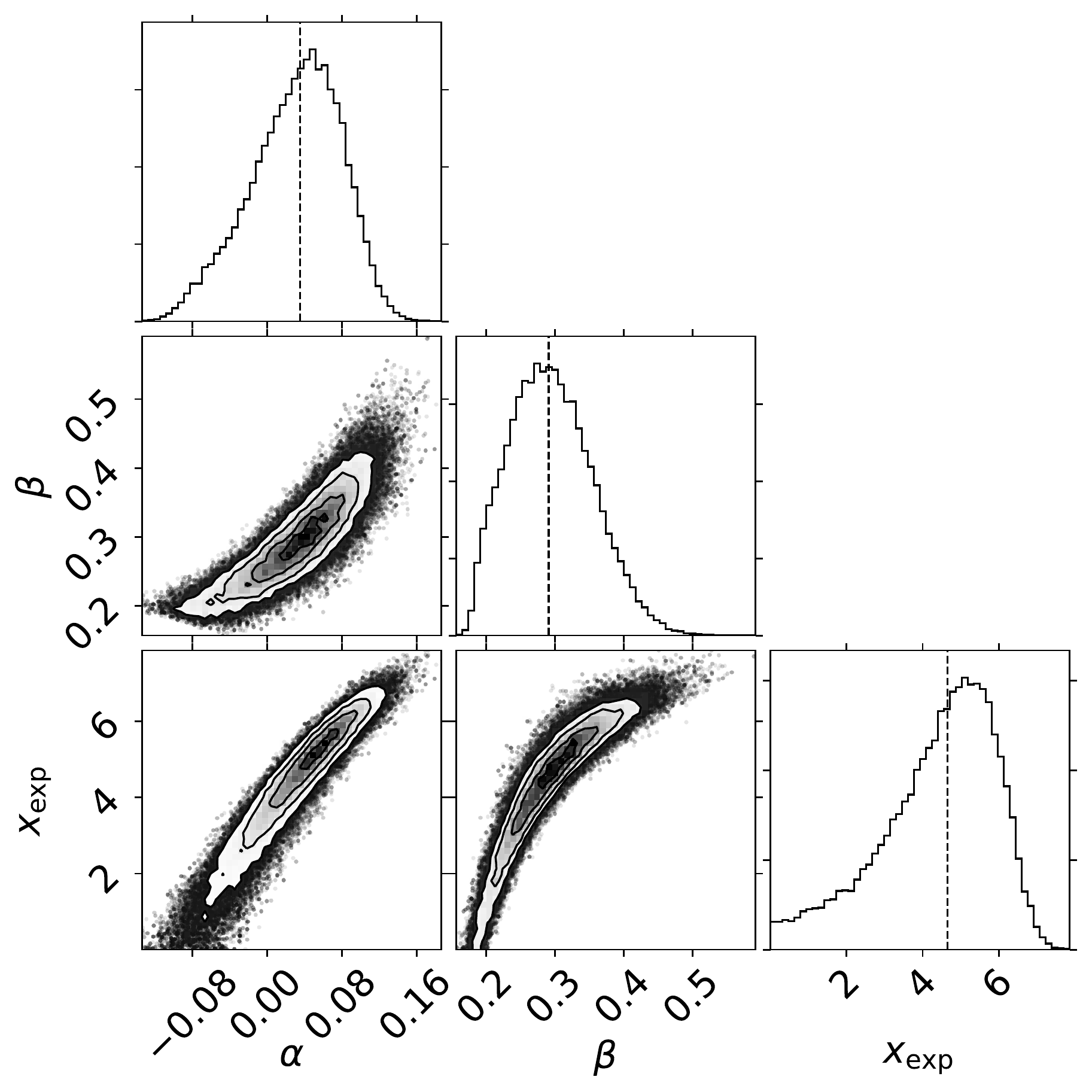}
  \caption{The posterior likelihood sampling results of fitting the Variably Tapered Power Law model (see Section \ref{subsec:models} and Table \ref{tab:models}) to the data from Run I using MCMC.  The median values are shown by the vertical dashed lines. \label{fig:app_egMCMC}}
\end{figure}

\begin{figure}
  \centering
  \includegraphics[width=\linewidth]{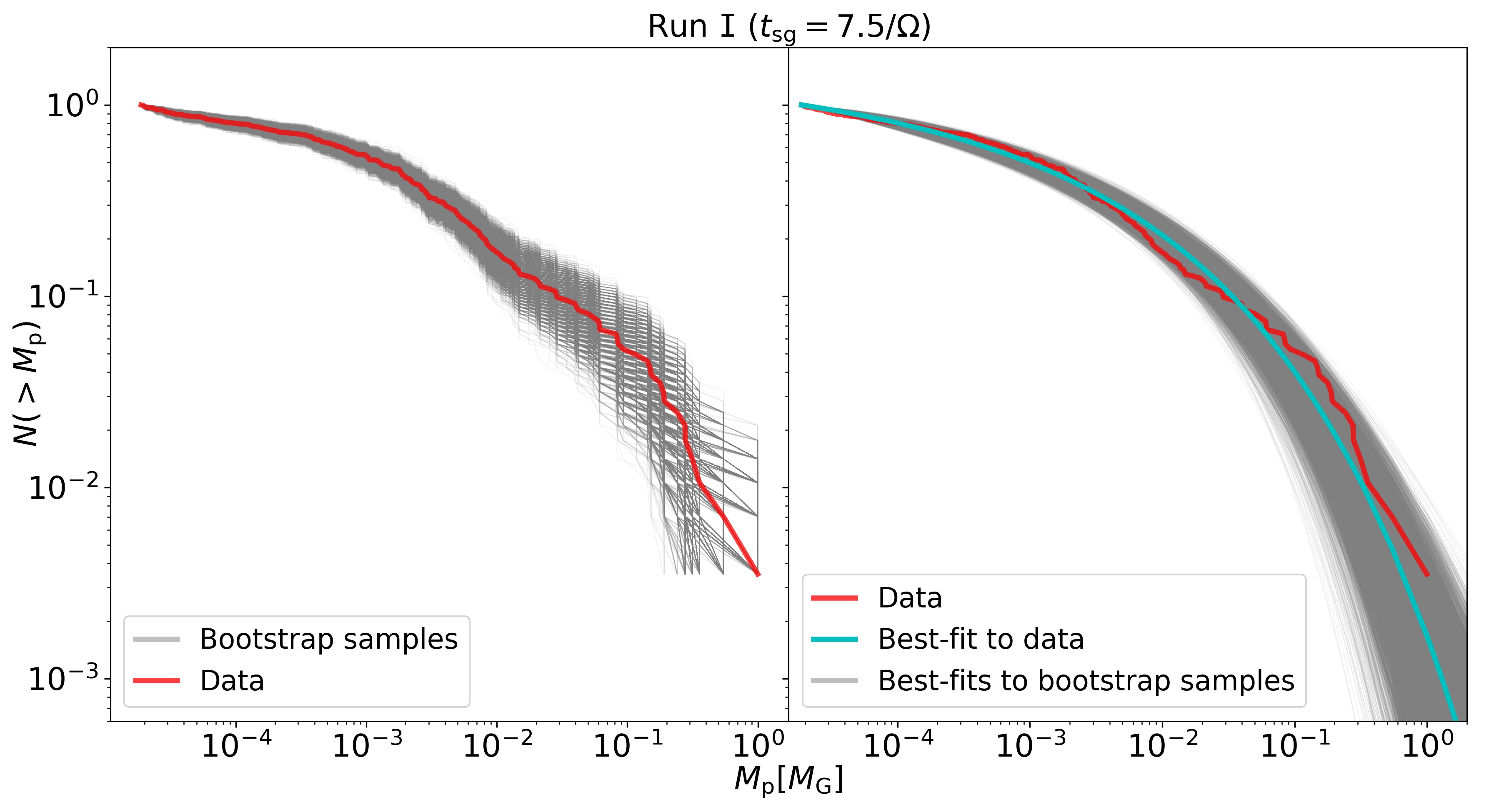}
  \caption{A demonstration of the nonparametric bootstrap method. \textit{left}: The simulated mass distribution data (red) and $N_{\rm bs}=10000$ bootstrap samples (grey). \textit{right:} The best-fit Variably Tapered Power Law model (cyan, $\bm{\theta}_{\rm MLE}$) to the data and the best-fit model to each bootstrap sample (grey, $\bm{\theta}_{\mathrm{bs}, k}$). \label{fig:app_egBMS}}
\end{figure}

In this appendix section, we take the fitting of the Variably Tapered Power Law model to the data from Run I as an example and describe it in detail.  First, we assume a uniform prior distribution of parameters and use \texttt{emcee} to explore the posterior likelihood sampling (see Fig. \ref{fig:app_egMCMC}).  In this example, our chains consist of 32 walkers and 6000 steps with 1000 burn-in steps, while the estimated autocorrelation time returned by \texttt{emcee} is only about 75 steps and is well blow the number of burn-in steps.  From this MCMC result, the best-fit parameters and the log-likelihood are
\begin{equation}
  \begin{aligned}
    \bm{\theta}_{\rm MCMC} &= (\alpha, \beta, x_{\rm exp}) = (0.035, 0.291, 4.657), \\
    -\ln\mathcal{L} &= 633.716.
  \end{aligned}
\end{equation}

We now feed a set of initial guesses generated on a mesh grid centered on the $\bm{\theta}_{\rm MCMC}$ to the \texttt{minimize} method provided by the \texttt{scipy.optimize} package.  Various minimization algorithms are then employed for further minimization, including ``Nelder-Mead'', ``Powell'', ``CG'', ``BFGS'', ``Newton-CG'', ``L-BFGS-B'' and ``TNC''.  The latter five algorithms compute the gradient vector, $\partial\ln\mathcal{L}/\partial\bm{\theta}$, to converge more quickly to the solution.  We have implemented the methods to compute the gradient vector and Hessian matrix for all seven statistical models and make them available on \href{https://github.com/astroboylrx/Pyridoxine/blob/master/pyridoxine/utility/stats.py}{GitHub}.  In this specific example, all algorithms but ``L-BFGS-B'' converge at the final best-fit parameters with an acceptable tiny gradient vector
\begin{equation}
  \begin{aligned}
    \bm{\theta}_{\rm MLE} &= (\alpha, \beta, x_{\rm exp}) = (0.036, 0.298, 4.734), \\
    \left.\left(\fracp{\ln\mathcal{L}}{\alpha}, \fracp{\ln\mathcal{L}}{\beta}, \fracp{\ln\mathcal{L}}{x_{\rm exp}} \right)\right|_{\bm{\theta}=\bm{\theta}_{\rm MLE}} &= (\num{3.763e-11}, \num{1.265e-11}, \num{-1.812e-12}), \\
    \left.\left(\fracpp{\ln\mathcal{L}}{\alpha}, \fracpp{\ln\mathcal{L}}{\beta}, \fracpp{\ln\mathcal{L}}{x_{\rm exp}} \right)\right|_{\bm{\theta}=\bm{\theta}_{\rm MLE}} &= (-6365.209, -3364.693, -18.445), \\
    -\ln\mathcal{L} &= 633.695,
  \end{aligned}
\end{equation}
where Eq. \ref{eq:ddL} is satisfied with a good numerical precision.

In the next step, we use the nonparametric bootstrap method to estimate the uncertainties of $\bm{\theta}_{\rm MLE}$.  Fig. \ref{fig:app_egBMS} shows all the bootstrap samples in the left panel.  We again apply our MLE to obtain the best-fit parameters, $\bm{\theta}_{\mathrm{bs}, k}$, for the $k$-th bootstrap sample ($k=1, \cdots, N_{\rm bs}$).  All these best-fits are plotted in the right panel of Fig. \ref{fig:app_egBMS}.  The uncertainties are then calculated based on Eq. \ref{eq:BMS_sigma}
\begin{equation}
  \begin{aligned}
    \alpha &= 0.036^{+0.041}_{-0.041}, \\
    \beta &= 0.298^{+0.061}_{-0.040}, \\
    x_{\rm exp} &= 4.734^{+1.022}_{-1.128}.
  \end{aligned}  
\end{equation}
Furthermore, the uncertainty of the characteristic mass scale, $M_{\rm exp}$, can be derived as
\begin{equation}
  \begin{aligned}
    M_{\rm exp} + \Delta M_{\rm exp}^+ = M_{\rm min} e^{x_{\rm exp}+\Delta x_{\rm exp}^+} &\Longrightarrow \Delta M_{\rm exp}^+ = M_{\rm exp} \left(e^{\Delta x_{\rm exp}^+} - 1 \right), \\
    M_{\rm exp} - \Delta M_{\rm exp}^- = M_{\rm min} e^{x_{\rm exp}-\Delta x_{\rm exp}^-} &\Longrightarrow \Delta M_{\rm exp}^- = M_{\rm exp} \left(1 - e^{-\Delta x_{\rm exp}^-} \right), \\
    \therefore M_{\rm exp} &= 0.0021^{+0.0038}_{-0.0014}.
  \end{aligned}
\end{equation}

\section{Model Coefficients and Full Functional Forms}
\label{appsec:model_coeff}

In this section, we list all the renormalization coefficients for the statistical models in Section \ref{subsec:models} and show their full functional forms in Table \ref{tab:models}.

\begin{enumerate}
  \item \textit{Simply Tapered Power Law}
    \begin{equation}
      c_1 = \frac{1}{M_{\rm min}^{-\alpha}}\ \exp\left(\frac{M_{\rm min}}{M_{\rm exp}}\right).
    \end{equation}
  \item \textit{Variably Tapered Power Law}
    \begin{equation}
      c_2 = \frac{1}{M_{\rm min}^{-\alpha}}\ \exp\left[\left(\frac{M_{\rm min}}{M_{\rm exp}}\right)^\beta\right].
    \end{equation}
  \item \textit{Broken Cumulative Power Law}
    \begin{equation}
      \begin{aligned}
        c_{31}&= \frac{1}{M_{\rm min}^{-\alpha_1}} \\
        c_{32}&= \frac{1}{M_{\rm min}^{-\alpha_1} M_{\rm br}^{\alpha_1-\alpha_2} }
      \end{aligned}
    \end{equation}
  \item \textit{Truncated Power Law}
    \begin{equation}
      c_4 = \frac{ \alpha }{M_{\rm min}^{-\alpha} - M_{\rm tr}^{-\alpha}}
    \end{equation}
  \item \textit{Broken Power Law}
    \begin{equation}
      \begin{aligned}
        c_{51}&= \frac{1}{M_{\rm min}^{-\alpha_1}} \left[\frac{1}{\alpha_1}+\left(\frac{1}{\alpha_2} - \frac{1}{\alpha_1} \right)\left(\frac{M_{\rm br}}{M_{\rm min}} \right)^{-\alpha_1} \right]^{-1} \\
        c_{52}&= c_{51} M_{\rm br}^{\alpha_2-\alpha_1}
      \end{aligned}
    \end{equation}
  \item \textit{Truncated Broken Power Law}
    \begin{equation}
      \begin{aligned}
        c_{61}&= \frac{1}{M_{\rm min}^{-\alpha_1}} \left[\frac{1}{\alpha_1}+\left(\frac{1}{\alpha_2} - \frac{1}{\alpha_1} \right)\left(\frac{M_{\rm br}}{M_{\rm min}} \right)^{-\alpha_1} - \frac{1}{\alpha_2} \left(\frac{M_{\rm br}}{M_{\rm min}} \right)^{\alpha_2-\alpha_1}\left(\frac{M_{\rm tr}}{M_{\rm min}} \right)^{-\alpha_2} \right]^{-1} \\
        c_{62}&= c_{61} M_{\rm br}^{\alpha_2-\alpha_1}
      \end{aligned}
    \end{equation}
  \item \textit{Three-segment Power Law}
    \begin{equation}
      \begin{aligned}
        c_{71}&= \frac{1}{M_{\rm min}^{-\alpha_1}} \left[\frac{1}{\alpha_1}+\left(\frac{1}{\alpha_2} - \frac{1}{\alpha_1} \right)\left(\frac{M_{\rm br1}}{M_{\rm min}} \right)^{-\alpha_1} + \left(\frac{1}{\alpha_3} - \frac{1}{\alpha_2} \right) \left(\frac{M_{\rm br1}}{M_{\rm min}} \right)^{\alpha_2-\alpha_1}\left(\frac{M_{\rm br2}}{M_{\rm min}} \right)^{-\alpha_2} \right]^{-1} \\
        c_{72}&= c_{71} M_{\rm br1}^{\alpha_2-\alpha_1} \\
        c_{73}&= c_{71} M_{\rm br1}^{\alpha_2-\alpha_1} M_{\rm br2}^{\alpha_3-\alpha_2}
      \end{aligned}
    \end{equation}
\end{enumerate}

\begin{deluxetable*}{c|c|c|c}%[!htbp]
  \tablecaption{Mass Distribution Models\label{tab:models}}
  \tabletypesize{\footnotesize}
  \tablecolumns{5}
  \tablewidth{0.0\linewidth}
  \tablehead{
    \colhead{Name} &
    \colhead{Mass Distribution Function} &
    \colhead{Mass Scale} &
    \colhead{PDF in Likelihood Estimator} \\
    %\cline{1-3}%\cline{3-3}
    \colhead{K: \# of parameters} &
    \colhead{CDF[$P_>(M)$] or PDF[$\xi(M)$]} &
    \colhead{$\displaystyle{x\equiv\ln{(M/M_{\rm min})}}$} &
    \colhead{$p(x;\bm{\theta})$}
  }
  \startdata
  \hline\hline
  \begin{minipage}[c][1.8cm][c]{0.16\textwidth}
    \centering Simply Tapered \\ Power Law \\ \citep{Abod2019} \\ K=2, $\bm{\theta}=(\alpha, x_{\rm exp})$
  \end{minipage}
  & $ \displaystyle{P_>(M)=\left(\frac{M}{M_{\rm min}}\right)^{-\alpha}\ \exp\left[-\frac{M-M_{\rm min}}{M_{\rm exp}}\right]} $
  & $ \displaystyle{x_{\rm exp}\equiv\ln{\left(\frac{M_{\rm exp}}{M_{\rm min}}\right)}} $
  & $ \displaystyle{\frac{ \left(\alpha + e^{-x_{\rm exp}}e^{x}\right) }{ \exp\left[\alpha x + e^{-x_{\rm exp}} (e^x-1)\right] } } $ \\
  \hline
  \begin{minipage}[c][1.8cm][c]{0.16\textwidth}
    \centering Variably Tapered \\ Power Law \\ \citep{Schafer2017} \\ K=3, $\bm{\theta}=(\alpha, \beta, x_{\rm exp})$
  \end{minipage}
  & $ \displaystyle{P_>(M)=\left(\frac{M}{M_{\rm min}}\right)^{-\alpha}\ \exp\left[-\frac{M^\beta-M_{\rm min}^\beta}{M_{\rm exp}^\beta}\right]} $
  & $ \displaystyle{x_{\rm exp}\equiv\ln{\left(\frac{M_{\rm exp}}{M_{\rm min}}\right)}} $
  & $ \displaystyle{\frac{ (\alpha + \beta e^{\beta (x - x_{\rm exp})}) }{ \exp\left[\alpha x + e^{-\beta x_{\rm exp}} (e^{\beta x}-1)\right] } }  $ \\
  \hline
  \begin{minipage}[c][2cm][c]{0.16\textwidth}
    \centering Broken Cumulative Power Law \\ K=3, $\bm{\theta}=(\alpha_1, \alpha_2, x_{\rm br})$
  \end{minipage}
  & $ P_>(M) = \left\{\begin{aligned}
        & \displaystyle{ \left(\frac{M}{M_{\rm min}}\right)^{-\alpha_1}}\ &\displaystyle{M\leqslant M_{\rm br}} \\
        & \displaystyle{ \frac{ M^{-\alpha_2} }{M_{\rm br}^{\alpha_1-\alpha_2} M_{\rm min}^{-\alpha_1}} }\ &\displaystyle{M>M_{\rm br}}
      \end{aligned}\right.$
  & $ \displaystyle{x_{\rm br}\equiv\ln{\left(\frac{M_{\rm br}}{M_{\rm min}}\right)}} $
  & $ \left\{\begin{aligned}
        & \alpha_1 e^{-\alpha_1 x}\ &x\leqslant x_{\rm br} \\
        & \alpha_2 e^{(\alpha_2 - \alpha_1)x_{\rm br} - \alpha_2 x}\ &x>x_{\rm br}
      \end{aligned}\right.$ \\
  \hline\hline
  \begin{minipage}[c][1.8cm][c]{0.16\textwidth}
    \centering Truncated Power Law \\ \citep{Schafer2017} \\ K=2, $\bm{\theta}=(\alpha, x_{\rm tr})$
  \end{minipage}
  & $ \displaystyle{\xi(M)} = \left\{\begin{aligned}
        & \displaystyle{\frac{\alpha}{M} \frac{ \left( M / M_{\rm min} \right)^{-\alpha} }{1 - \left( M_{\rm tr}/M_{\rm min} \right)^{-\alpha}} }\ &\displaystyle{M\leqslant M_{\rm tr}} \\
        & 0\ &\displaystyle{M>M_{\rm tr}}
      \end{aligned}\right. $
  & $ \displaystyle{x_{\rm tr}\equiv\ln{\left(\frac{M_{\rm tr}}{M_{\rm min}}\right)}}  $
  & $ \left\{\begin{aligned}
        & \displaystyle{\frac{\alpha e^{-\alpha x}}{  1 - e^{-\alpha x_{\rm tr}} }}\ &\displaystyle{x\leqslant x_{\rm tr}} \\
        & 0\ &\displaystyle{x>x_{\rm tr}}
      \end{aligned}\right. $ \\
  \hline
  \begin{minipage}[c][3.25cm][c]{0.16\textwidth}
    \centering Broken Power Law \\ K=3, $\bm{\theta}=(\alpha_1, \alpha_2, x_{\rm br})$
  \end{minipage}
  & $ \begin{aligned}
        \displaystyle{\xi(M)} = \left\{\begin{aligned}
          & \displaystyle{\frac{C_0}{M} \left(\frac{M}{M_{\rm min}}\right)^{-\alpha_1} }\ &\displaystyle{M\leqslant M_{\rm br}} \\
          & \displaystyle{\frac{C_0}{M} \frac{ \left(M/M_{\rm min}\right)^{-\alpha_2} }{ \left(M_{\rm br}/M_{\rm min}\right)^{\alpha_1-\alpha_2} }}\ &\displaystyle{M>M_{\rm br}}
        \end{aligned}\right. & \\
        \displaystyle{ C_0 = \left[\frac{1}{\alpha_1}+\left(\frac{1}{\alpha_2} - \frac{1}{\alpha_1} \right)\left(\frac{M_{\rm br}}{M_{\rm min}} \right)^{-\alpha_1} \right]^{-1} } &
      \end{aligned} $
  & $ \displaystyle{x_{\rm br}\equiv\ln{\left(\frac{M_{\rm br}}{M_{\rm min}}\right)}} $
  & $ \begin{aligned}
        \left\{\begin{aligned}
          & \displaystyle{C_0 e^{-\alpha_1 x}}\ &\displaystyle{x\leqslant x_{\rm br}} \\
          & \displaystyle{C_0 e^{(\alpha_2 - \alpha_1)x_{\rm br} -\alpha_2 x}}\ &\displaystyle{x>x_{\rm br}}
        \end{aligned}\right.\ & \\
        \displaystyle{C_0 = \left[ \frac{1}{\alpha_1}+\left(\frac{1}{\alpha_2} - \frac{1}{\alpha_1} \right) e^{-\alpha_1 x_{\rm br}} \right]^{-1} } &
      \end{aligned} $ \\
  \hline
  \begin{minipage}[c][4.5cm][c]{0.16\textwidth}
    \centering Truncated Broken Power Law \\ K=4, $\bm{\theta}=(\alpha_1, \alpha_2, x_{\rm br}, x_{\rm tr})$
  \end{minipage}
  & $ \begin{aligned}
        \displaystyle{\xi(M)} = \left\{\begin{aligned}
          & \displaystyle{\frac{C_1}{M} \left(\frac{M}{M_{\rm min}}\right)^{-\alpha_1} }\ &\displaystyle{M\leqslant M_{\rm br}} \\
          & \displaystyle{\frac{C_1}{M} \frac{ \left(M/M_{\rm min}\right)^{-\alpha_2} }{ \left(M_{\rm br}/M_{\rm min}\right)^{\alpha_1-\alpha_2} }}\ &\text{otherwise} \\
          & 0\ &\displaystyle{M>M_{\rm tr}}
        \end{aligned}\right. & \\
        \pushright{ \displaystyle{ C_1 = \left[ \frac{1}{\alpha_1} + \left(\frac{1}{\alpha_2} - \frac{1}{\alpha_1} \right) \left(\frac{M_{\rm br}}{M_{\rm min}} \right)^{-\alpha_1} \right. }} & \\
        \pushright{ \displaystyle{ \left. - \frac{1}{\alpha_2} \left(\frac{M_{\rm br}}{M_{\rm min}} \right)^{\alpha_2-\alpha_1}\left(\frac{M_{\rm tr}}{M_{\rm min}} \right)^{-\alpha_2}\right]^{-1} }} &
      \end{aligned} $
  & $ \begin{aligned}
    & \displaystyle{x_{\rm br}\equiv\ln{\left(\frac{M_{\rm br}}{M_{\rm min}}\right)}} \\
    & \displaystyle{x_{\rm tr}\equiv\ln{\left(\frac{M_{\rm tr}}{M_{\rm min}}\right)}}
  \end{aligned} $
  & $ \begin{aligned}
        \left\{\begin{aligned}
          & \displaystyle{C_1 e^{-\alpha_1 x}}\ &\displaystyle{x\leqslant x_{\rm br}} \\
          & \displaystyle{C_1 e^{(\alpha_2 - \alpha_1)x_{\rm br} -\alpha_2 x}}\ &\text{otherwise} \\
          & 0\ &\displaystyle{x>x_{\rm tr}}
        \end{aligned}\right.\ & \\
        \displaystyle{C_1 = \left[ \frac{1}{\alpha_1} + \left(\frac{1}{\alpha_2} - \frac{1}{\alpha_1} \right) e^{-\alpha_1 x_{\rm br}} \right.} & \\
        \displaystyle{\left. - \frac{1}{\alpha_2} e^{(\alpha_2-\alpha_1) x_{\rm br} - \alpha_2 x_{\rm tr}} \right]^{-1} } &
      \end{aligned} $ \\
  \hline
  \begin{minipage}[c][5.75cm][c]{0.16\textwidth}
    \centering Three-segment \\ Power Law \\ K=5, $\bm{\theta}=(\alpha_1, \alpha_2, \alpha_3, x_{\rm br1}, x_{\rm br2})$
  \end{minipage}
    & $ \begin{aligned}
          & \displaystyle{\xi(M)} = \hfill &\\
          & \quad \left\{\begin{aligned}
              & \displaystyle{\frac{C_2}{M} \left(\frac{M}{M_{\rm min}}\right)^{-\alpha_1}}\ &\displaystyle{M\leqslant M_{\rm br1}} \\
              & \displaystyle{\frac{C_2}{M} \frac{ \left(M/M_{\rm min}\right)^{-\alpha_2} }{ \left(M_{\rm br1}/M_{\rm min}\right)^{\alpha_1-\alpha_2} }}\ &\text{otherwise} \\
              & \displaystyle{\frac{C_2}{M} \frac{ \left(M/M_{\rm min}\right)^{-\alpha_3} }{ \displaystyle{\left(\frac{M_{\rm br1}}{M_{\rm min}}\right)^{\alpha_1-\alpha_2} \left(\frac{M_{\rm br2}}{M_{\rm min}}\right)^{\alpha_2-\alpha_3}} }}\ &\displaystyle{M>M_{\rm br2}}
            \end{aligned}\right. & \\
          & \pushright{ \displaystyle{ C_2 = \left[ \frac{1}{\alpha_1} + \left(\frac{1}{\alpha_2} - \frac{1}{\alpha_1} \right) \left(\frac{M_{\rm br1}}{M_{\rm min}} \right)^{-\alpha_1} \right. }} & \\
          & \pushright{ \displaystyle{ \left. + \left(\frac{1}{\alpha_3} - \frac{1}{\alpha_2} \right) \left(\frac{M_{\rm br1}}{M_{\rm min}} \right)^{\alpha_2-\alpha_1}\left(\frac{M_{\rm br2}}{M_{\rm min}} \right)^{-\alpha_2}\right]^{-1} }} &
        \end{aligned}
      $
  & $ \begin{aligned}
        & \displaystyle{x_{\rm br1}\equiv\ln{\left(\frac{M_{\rm br1}}{M_{\rm min}}\right)}} \\
        & \displaystyle{x_{\rm br2}\equiv\ln{\left(\frac{M_{\rm br2}}{M_{\rm min}}\right)}}
      \end{aligned} $
  & $ \begin{aligned}
        \left\{\begin{aligned}
          & \displaystyle{ C_2 e^{-\alpha_1 x}}\ &x\leqslant x_{\rm br1} \\
          & \displaystyle{ C_2 e^{(\alpha_2 - \alpha_1)x_{\rm br1} -\alpha_2 x}}\ &\text{otherwise} \\
          & \displaystyle{ C_2 \frac{e^{-\alpha_3 x}}{ e^{(\alpha_1 - \alpha_2)x_{\rm br1} + (\alpha_2 - \alpha_3)x_{\rm br2}}} }\ &x>x_{\rm br2} \\
        \end{aligned}\right. & \\
        \displaystyle{C_2 = \left[ \frac{1}{\alpha_1} + \left(\frac{1}{\alpha_2} - \frac{1}{\alpha_1} \right) e^{-\alpha_1 x_{\rm br1}} \right.} & \\
        \displaystyle{\left. + \left(\frac{1}{\alpha_3} - \frac{1}{\alpha_2} \right) e^{(\alpha_2-\alpha_1) x_{\rm br1} - \alpha_2 x_{\rm br2}} \right]^{-1} } &
      \end{aligned} $ \\
  \enddata
  \tablecomments{In this table, $\alpha$ indicates the power law indices and $\beta$ denotes tapering indices.}
  %\tablenotetext{*}{N.B., $\displaystyle{ p(x;\bm{\theta}) = \frac{dn}{dx} = \frac{dn}{d\ln{M}} }$}
\end{deluxetable*}

%%%%%%%%%%%%%%%%%%%% REFERENCES %%%%%%%%%%%%%%%%%%

% The best way to enter references is to use BibTeX:

\bibliographystyle{aasjournal}
\bibliography{refs}

%%%%%%%%%%%%%%%%%%%%%%%%%%%%%%%%%%%%%%%%%%%%%%%%%%

% \listofchanges % does not work properly

%%%%%%%%%%%%%%%%%%%%%%%%%%%%%%%%%%%%%%%%%%%%%%%%%%
\end{document}